\documentclass[seceq]{ptptex}

\usepackage{graphicx}
\usepackage{wrapft}

\title{
Theory of successive transitions in vanadium spinels \\
and order of orbitals and spins
}

\author{
Yukitoshi \textsc{Motome}$^1$ and Hirokazu \textsc{Tsunetsugu}$^2$
}

\inst{
${}^1$RIKEN (The Institute of Physical and Chemical Research), 
Saitama 351-0198 \\
${}^2$Yukawa Institute for Theoretical Physics, 
Kyoto University, Kyoto 606-8502
}

\abst{
We have theoretically studied successive transitions 
in vanadium spinel oxides with ($t_{2g}$)$^2$ electron configuration. 
These compounds show a structural transition at $\sim 50$K and 
an antiferromagnetic transition at $\sim 40$K. 
Since threefold $t_{2g}$ orbitals of vanadium cations are occupied partially 
and vanadiums constitute a geometrically--frustrated pyrochlore lattice, 
the system provides a particular example to investigate 
the interplay among spin, orbital and lattice degrees of freedom 
on frustrated lattice. 
We examine the models with the Jahn--Teller coupling and/or 
the spin--orbital superexchange interaction, 
and conclude that keen competition between these two contributions 
explains the thermodynamics of vanadium spinels. 
Effects of quantum fluctuations as well as relativistic 
spin--orbit coupling are also discussed.
}

\begin{document}

\maketitle


\section{Introduction}
\label{sec:intro}

\begin{figure}[bt]
\centerline{\includegraphics[width=9cm]{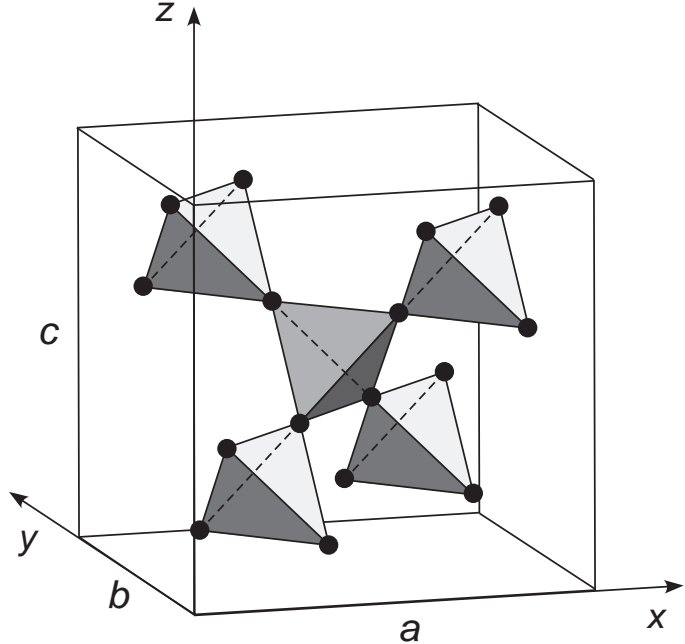}}
\caption{
Cubic unit cell of the pyrochlore lattice. 
In vanadium spinels, vanadium cations constitute 
this pyrochlore structure.
}
\label{fig:pyrochlore}
\end{figure}

In this paper, we review our recent studies 
on phase transitions in vanadium spinel compounds, 
$A$V$_2$O$_4$ ($A$=Zn, Mg or Cd). 
\cite{Tsunetsugu2003,Motome2004,Motome2005a,Motome2005b,MotomePREPRINT} 
This family of materials is antiferromagnetic insulator with 
spin $S=1$ on pyrochlore sublattice in the spinel structure
(Fig.~\ref{fig:pyrochlore}). 
Recent experiments showed the presence of two 
successive phase transitions at low temperatures around $50$K.  
\cite{Ueda1997}
As temperature is lowered, a structural transition 
occurs first, then a magnetic transition follows. 
As we will discuss later, simple spin 
exchange mechanism does not explain the lowest 
temperature phase, and we need to take account 
of two important factors in order to explain 
basic character of these transitions.  
One is geometrical frustration and the other 
is strong correlation effects, which lead to 
keen competition of various degrees of freedom.  
As widely known, the geometrical frustration often leads to 
huge degeneracy of the ground--state manifold, and 
suppresses a long--range ordering. 
On the other hand, the strong electron correlation opens
several channels in the low energy sector, such as
spin and orbital degrees of freedom.
A key question is how these channels affect 
the manifold --- do they result in some nontrivial
phenomena through enhanced fluctuations, or 
lift the degeneracy in some nontrivial manner?
We would like to clarify this intriguing problem 
by studying effects of both frustration and correlation
in the vanadium spinels.

Vanadium spinels are a $3d$--electron system with 
orbital degrees of freedom.   
In the last decade, various 
transition--metal materials have been intensively 
studied to investigate the effects of orbital 
degrees of freedom.  
\cite{Tokura2000}
In particular, manganites showing colossal 
magnetoresistance (CMR) are the most typical 
example.  Vanadium spinels have different features 
from these materials.  
A distinct difference is the lattice structure.
In manganites showing CMR, the lattice structure is perovskite;
atomic bonds are connected with right angle 
and the network of magnetic ions is close to a simple cubic lattice. 
In spinels, atomic bonds are connected to constitute 
a lattice in which elementary plaquettes are triangles;  
the network of magnetic cations is the pyrochlore lattice
shown in Fig.~\ref{fig:pyrochlore}.
Another important difference is that the vanadates are 
a $t_{2g}$--electron system, whereas the manganites have 
conduction electrons in $e_g$ multiplet.  
Jahn--Teller distortions are generally smaller in $t_{2g}$ systems, and 
the crystal--field splitting of the electronic levels is less prominent, 
partly because $t_{2g}$ orbitals point away from surrounding anions. 
Therefore, it is important to take account of orbital degrees of freedom 
in $t_{2g}$ systems like vanadium spinels. 
We also note that $t_{2g}$ electrons may have a
nonzero orbital angular momentum, whereas orbital 
angular momenta are quenched in $e_g$ systems.  
This also leads to several effects present 
only in $t_{2g}$ systems.  For example, relativistic 
spin--orbit coupling is sometimes important to 
take into account, but this is not a central 
issue in this paper.

\begin{figure}[bt]
\centerline{\includegraphics[width=9cm]{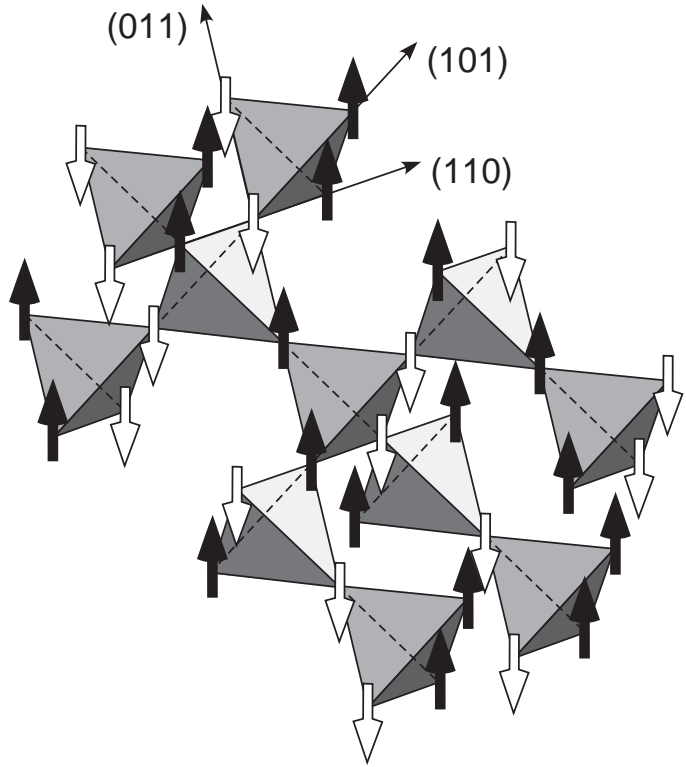}}
\caption{
Magnetic ordering structure in the lowest--temperature phase 
of vanadium spinels 
determined by neutron scattering experiment. 
\cite{Niziol1973}
}
\label{fig:neutron}
\end{figure}

From the viewpoints of orbital physics and geometrical 
frustration, spinels with transition--metal ions 
are very interesting system.  In this paper, we shall 
discuss about two phase transitions in vanadium 
spinels with divalent $A$--site cations 
such as Zn, Mg or Cd.
In these materials, first of all, 
interesting phenomena such as phase transitions occur
at much lower temperatures compared to a typical energy scale 
in the system.
That is, the characteristic energy scale, which is given by, 
for example, the Curie--Weiss temperature estimated from 
the temperature dependence of the magnetic susceptibility, 
is $\sim 1000$K,
\cite{Muhtar1988}
nevertheless the system remains paramagnetic and does not 
show any transition down to $\sim 50$K. 
This suppression of relevant energy scale is 
characteristic to frustrated systems. 
At low temperatures, in the end, the compounds show two successive transitions.
\cite{Ueda1997}
One is a structural transition at $\sim 50$K 
and the lattice changes its symmetry. 
While the lattice symmetry is cubic above the transition temperature, 
it changes to tetragonal below the temperature, and 
the system is compressed 
along one of the principal axes, the $c$ axis. 
The other transition takes place at a lower temperature $\sim 40$K,
and ascribed to an antiferromagnetic ordering. 
The magnetic structure in the lowest--temperature phase was 
determined by neutron scattering experiment 
as shown in Fig.~\ref{fig:neutron}. 
\cite{Niziol1973}
It consists of up--down--up--down-- staggered antiferromagnetic pattern
along the chains in the $xy$ planes [($110$) and ($1\bar10$) directions]
and up--up--down--down-- pattern with period four
along the chains in the $yz$ and $zx$ planes 
[($101$), ($10\bar1$), ($011$) and ($01\bar1$) directions]. 
Thus the fundamental questions are 
(i) what the driving mechanism of these successive transitions is and
(ii) how the complicated magnetic ordering is stabilized. 
We will answer these questions theoretically 
by investigating keen competitions among spin, orbital and lattice
degrees of freedom in this geometrically--frustrated system.

This paper is organized as follows. 
In Sec.~\ref{sec:Jahn-Teller}, we examine a model with 
the Jahn--Teller coupling alone.
We also examine a model with spin--orbital superexchange interaction alone 
in Sec.~\ref{sec:superexchange}. 
In these two sections, we show that both two models fail to explain 
the experimental results. 
In Sec.~\ref{sec:superexchange+Jahn-Teller},
we take into account both two terms in the model and 
study keen competition between them. 
We show, by using mean--field analysis and Monte Carlo simulation, that 
the model successfully explains two successive transitions 
of vanadium spinels. 
Effects of fluctuations, both thermal and quantum, are also discussed. 
In Sec.~\ref{sec:disc}, we make some remarks on 
effects of relativistic spin--orbital coupling, 
direction of magnetic moment, and 
a possibility of Haldane gap state in this system. 
Section \ref{sec:summary} is devoted to summary.

\section{Model with Jahn--Teller coupling}
\label{sec:Jahn-Teller}

\begin{figure}[bt]
\centerline{\includegraphics[width=14cm]{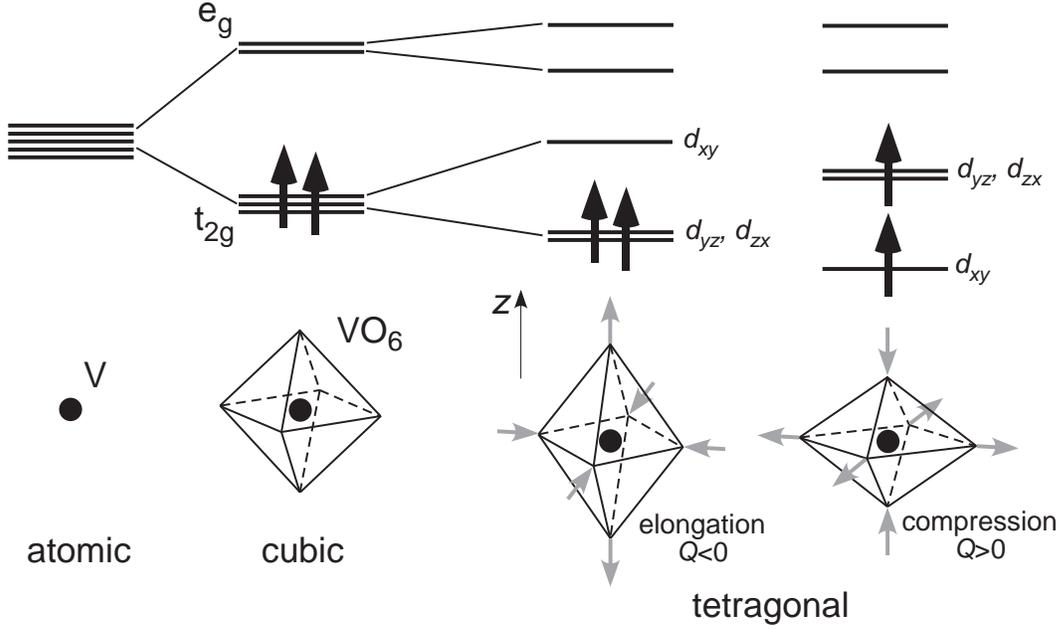}}
\caption{
Crystal--field splitting of $d$ levels.
}
\label{fig:cf}
\end{figure}

Let us start from a simple model with the Jahn--Teller coupling alone, 
which describes a coupling of orbital degrees of freedom 
with lattice distortions. 
In the octahedral position surrounded by a symmetric O$_6$ octahedron, 
the cubic crystal field splits
the fivefold $d$ levels of vanadium cation into 
higher twofold $e_g$ levels and lower threefold $t_{2g}$ levels 
(left two columns in Fig.~\ref{fig:cf}).
Jahn--Teller distortion is a distortion of O$_6$ octahedron, 
which gains energy by lifting the degeneracy further 
to achieve a closed--shell configuration of electronic state. 
In other words, the Jahn--Teller effect is active only for the case of 
an open--shell configuration, and hence, depends on the number of $d$ electrons.

In vanadium spinels with divalent $A$--site cation $A$V$_2$O$_4$, 
each vanadium has valence V$^{3+}$, and 
this corresponds to ($t_{2g}$)$^2$ configuration. 
Hence, threefold $t_{2g}$ levels are partially filled 
(second column in Fig.~\ref{fig:cf}) 
so that the Jahn--Teller effect is active.
We start from an ionic model describing independent VO$_6$ octahedra, and 
later we will introduce intersite correlations.
If we consider the Jahn--Teller distortion of tetragonal symmetry,
which is observed in experiments,
the ionic model can be given in the form
\begin{equation}
H_{\rm JT}^{\rm ionic} = \sum_i
\gamma Q_i (n_{i1} + n_{i2} - 2 n_{i3}) + \frac12 \sum_i Q_i^2,
\label{eq:H_JT^ionic}
\end{equation}
where $\gamma$ is the electron--phonon coupling constant and 
$Q_i$ denotes the amplitude of local lattice distortion at site $i$. 
Here, we consider the tetragonal distortion along one of the principal axes, 
the $z$ direction, and
choose the sign of $Q_i$ such that it is positive for a compression 
of the octahedron in the $z$ direction. 
The orbital indices $\alpha = 1,2$ and $3$ of the density operator $n_{i\alpha}$ 
denotes $d_{yz}, d_{zx}$ and $d_{xy}$, respectively.
Assuming high--spin states by strong Hund's--rule coupling, 
the model is defined for spinless fermions.
The second term describes the elastic energy 
($Q$ is normalized to absorb the elastic constant);
here we use the adiabatic approximation and 
phonons are regarded as classical objects.
The model (\ref{eq:H_JT^ionic}) is considered with the local constraint 
on the density as
\begin{equation}
\sum_\alpha \langle n_{i\alpha} \rangle = 2. 
\label{eq:constraints}
\end{equation}

The ground state of the ionic model~(\ref{eq:H_JT^ionic}) is obtained 
by minimizing the energy under the constraint~(\ref{eq:constraints}). 
The ground--state energy per site is given by
\begin{equation}
E_{\rm JT}^{\rm ionic} = - 2 \gamma^2 \quad {\rm for} \ \ 
Q_i = -2\gamma, 
\label{eq:Emin1^ionic}
\end{equation}
and for the orbital states
\begin{equation}
\langle n_{i1} \rangle = \langle n_{i2} \rangle = 1 \quad {\rm and} \quad
\langle n_{i3} \rangle = 0. 
\label{eq:sol1a}
\end{equation}
This situation corresponds to the elongation of VO$_6$ octahedra
as shown in the third column in Fig.~\ref{fig:cf}:
The doublet of $d_{yz}$ and $d_{zx}$ levels is lowered and 
occupied by two electrons to form the closed--shell configuration. 

Experimentally, lattice distortion occurs in the opposite way 
at low temperatures, 
that is, a uniform compression of octahedra. 
If we restrict to the case of compression $Q > 0$, 
the energy minimum is found as
\begin{equation}
\tilde{E}_{\rm JT}^{\rm ionic} = - \frac{\gamma^2}{2} \quad {\rm for} \ \ 
Q_i = \gamma,
\label{eq:Emin2^ionic}
\end{equation}
and for the orbital states
\begin{equation}
\langle n_{i1} + n_{i2} \rangle = 1 \quad {\rm and} \quad
\langle n_{i3} \rangle = 1, 
\label{eq:sol2a}
\end{equation} 
which are shown in the last column in Fig.~\ref{fig:cf}.

\begin{figure}[bt]
\centerline{\includegraphics[width=7cm]{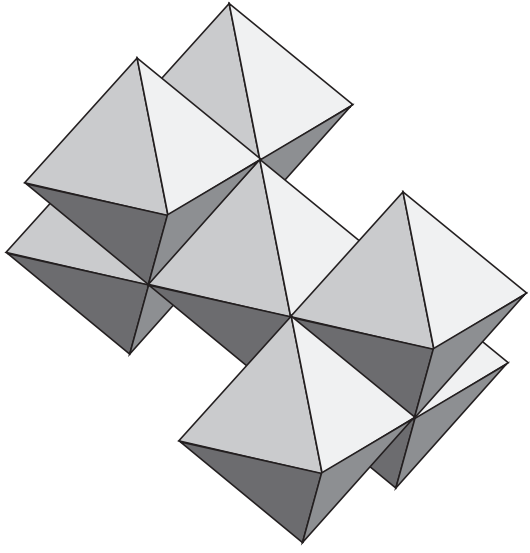}}
\caption{
Edge--sharing network of VO$_6$ octahedra in the spinel structure. 
Vanadium cations locate in the center of each octahedron and 
constitute the pyrochlore lattice in Fig.~\ref{fig:pyrochlore}. 
}
\label{fig:octahedra}
\end{figure}

In the real lattice structure, the Jahn--Teller effect should have 
some cooperative aspect 
--- VO$_6$ octahedra share oxygen ions between neighbors, 
which leads to intersite correlations of the Jahn--Teller distortion.
Since spinel lattice structure consists of three--dimensional network of 
edge--sharing octahedra as shown in Fig.~\ref{fig:octahedra}, 
a distortion of VO$_6$ octahedron tends to distort 
neighboring octahedra in the same manner.
In other words, a ferro--type correlation 
is expected for the tetragonal distortions between neighboring octahedra.
This tendency is mimicked by the ferro--type intersite coupling of the distortion as
\begin{equation}
H_{\rm JT}^{\rm nn} = - \lambda \sum_{\langle ij \rangle} Q_i Q_j,
\label{eq:H_JT^intersite}
\end{equation}
where $\lambda > 0$ is the coupling constant and 
the summation is taken over the nearest--neighbor sites
on the pyrochlore lattice which consists of vanadium cations.

If we omit farther--neighbor correlations beyond the nearest neighbors, 
the total Hamiltonian for the Jahn--Teller effect is given by
\begin{equation}
H_{\rm JT} = H_{\rm JT}^{\rm ionic} + H_{\rm JT}^{\rm nn}.
\label{eq:H_JT}
\end{equation}
We can minimize the total energy with respect to $Q_i$ 
by assuming uniform orbital states at all sites. 
This corresponds to find a mean--field solution for 
cooperative Jahn--Teller distortion of ferro--type,
which is naturally expected for the ferro--type intersite coupling 
in Eq.~(\ref{eq:H_JT^intersite}).
For the orbital state (\ref{eq:sol1a}), or equivalently 
for an elongation of all VO$_6$ octahedra, we obtain 
the minimum of energy per site as 
\begin{equation}
E_{\rm JT} = - \frac{2 \gamma^2}{1-6\lambda} \quad {\rm for} \ \ 
Q_i = - \frac{2 \gamma}{1-6\lambda}.
\label{eq:Emin1}
\end{equation}
Note that $0 < \lambda < 1/6$ is required to keep the system stable.
On the other hand, for the orbital state (\ref{eq:sol2a}), 
corresponding to a compression of octahedra, we find a minimum as
\begin{equation}
\tilde{E}_{\rm JT} = - \frac{\gamma^2}{2(1-6\lambda)} \quad {\rm for} \ \ 
Q_i = \frac{\gamma}{1-6\lambda}.
\label{eq:Emin2}
\end{equation}
Since $E_{\rm JT} < \tilde{E}_{\rm JT}$, 
the model with Jahn--Teller effect alone predicts 
a uniform elongation of all VO$_6$ octahedra, 
which apparently disagrees with the experimental result.
In order to stabilize the compression of octahedra 
as observed in experiments,
we need to compensate the energy difference per site
\begin{equation}
\Delta_{\rm JT} = \tilde{E}_{\rm JT} - E_{\rm JT} 
= \frac{3 \gamma^2}{2(1-6\lambda)}
\end{equation}
by some energy gain from other mechanisms.
We will see in Sec.~\ref{sec:superexchange+Jahn-Teller} 
that this compensation indeed comes from 
the orbital superexchange interaction.

\section{Model with superexchange interaction}
\label{sec:superexchange}

\subsection{Model Hamiltonian}
\label{sec:SEmodel}

Next we examine effects of spin and orbital superexchange interactions. 
The superexchange interactions describe effective intersite couplings 
of both spin and orbital degrees of freedom 
in low--energy physics of strongly-correlated insulators. 
They are derived from a multi--orbital Hubbard model 
by using the perturbation in the strong--correlation limit. 

We start from the Hubbard model with threefold orbital degeneracy 
corresponding to the $t_{2g}$ manifold. 
The Hamiltonian reads
\begin{equation}
H_{\rm Hub} = \sum_{ij} \sum_{\alpha \beta} \sum_\tau
(t_{i\alpha, j\beta} c_{i \alpha \tau}^\dagger c_{j \beta \tau} + {\rm H.c.})
+\frac12 \sum_i \sum_{\alpha\beta,\alpha'\beta'} \sum_{\tau\tau'} 
U_{\alpha\beta,\alpha'\beta'}
c_{i\alpha\tau}^{\dagger} c_{i\beta\tau'}^{\dagger}
c_{i\beta'\tau'} c_{i\alpha'\tau},
\label{eq:H_multiorbital}
\end{equation}
where $i,j$ and $\tau,\tau'$ are site and spin indices, respectively, 
and $\alpha,\beta = 1$ ($d_{yz}$), $2$ ($d_{zx}$), $3$ ($d_{xy}$) are orbital indices. 
The first term of $H_{\rm Hub}$ is the electron hopping, and 
the second term describes onsite Coulomb interactions, 
for which we use the standard parametrizations,
\begin{eqnarray}
U_{\alpha\beta,\alpha'\beta'} &=& U' \delta_{\alpha\alpha'} \delta_{\beta\beta'} 
+ J_{\rm H} (\delta_{\alpha\beta'} \delta_{\beta\alpha'} +
\delta_{\alpha\beta} \delta_{\alpha'\beta'}), 
\\
U &=& U' + 2J_{\rm H}.
\end{eqnarray}

In order to describe the low--energy effective theory 
for the insulating state of vanadium spinels, 
we apply the second--order perturbation in the hopping term. 
The unperturbed states are atomic eigenstates with two electrons 
on each vanadium cations in a high--spin state ($S=1$). 
Among various hopping processes, 
the most relevant contribution is the 
nearest-neighbor $\sigma$ bond, 
because the spinel lattice structure consists of 
three--dimensional network of edge--sharing VO$_6$ octahedra
in Fig.~\ref{fig:octahedra}. 
\cite{Harrison,Matsuno1999} 
This is the hopping in the case where one of four lobes 
of each $t_{2g}$ orbital is pointing towards the other end of the bond.
In the perturbation, we take account of this $\sigma$--bond contribution only
for simplicity. 
The superexchange Hamiltonian, 
which is called the Kugel--Khomskii type Hamiltonian, 
\cite{Kugel1973}
is obtained for the nearest--neighbor sites in the form
\begin{eqnarray}
&& H_{\rm SE}^{\rm nn} = -J \sum_{\langle ij \rangle} \
[ \ h_{\rm o-AF}^{(ij)} + h_{\rm o-F}^{(ij)} \ ],
\label{eq:H_SE^nn}
\\
&& h_{\rm o-AF}^{(ij)} = (A + B \mib{S}_i \cdot \mib{S}_j)
\big[ n_{i \alpha(ij)} (1 - n_{j \alpha(ij)}) 
+ (1 - n_{i \alpha(ij)}) n_{j \alpha(ij)} \big],
\label{eq:h_o-AF}
\\
&& h_{\rm o-F}^{(ij)} = C (1 - \mib{S}_i \cdot \mib{S}_j)
n_{i \alpha(ij)} n_{j \alpha(ij)}, 
\label{eq:h_o-F}
\end{eqnarray}
where $\mib{S}_i$ is the $S=1$ spin operator 
and $n_{i \alpha} = \sum_{\tau} c_{i \alpha \tau}^\dagger c_{i \alpha \tau}$ 
is the density operator for site $i$ and orbital $\alpha$. 
Here, $\alpha(ij)$ is the orbital which has a finite hopping integral
between the sites $i$ and $j$, for instance, 
$\alpha(ij) = 3$ ($d_{xy}$) for $i$ and $j$ sites in the same $xy$ plane. 
Parameters in Eqs.~(\ref{eq:H_SE^nn})--(\ref{eq:h_o-F}) are given  
by coupling constants in Eq.~(\ref{eq:H_multiorbital}) as 
\begin{eqnarray}
&& J = \frac{(t_{\sigma}^{\rm nn})^2}{U},
\label{eq:J}
\\
&& A = \frac{1-\eta}{1-3\eta}, \quad 
B = \frac{\eta}{1-3\eta}, \quad
C = \frac{1+\eta}{1+2\eta}, 
\label{eq:ABC}
\end{eqnarray}
where $t_{\sigma}^{\rm nn}$ is the hopping integral of $\sigma$ bond, 
and 
$\eta = J_{\rm H}/U$
is a small parameter of $\sim 0.1$. 
Note that each site is subject to the local constraint, 
$\sum_{\alpha=1}^3 n_{i \alpha} = 2$. 

The superexchange model (\ref{eq:H_SE^nn}) consists of two contributions, 
$h_{\rm o-AF}^{(ij)}$ and $h_{\rm o-F}^{(ij)}$: 
The former favors spin--ferro and orbital--antiferro configuration 
on $ij$ bond while
the latter favors spin--antiferro and orbital--ferro configuration. 
The competition between these two exclusive contributions is 
a general aspect of Kugel--Khomskii type models. 
Another characteristic feature is 
the difference in the symmetry of the interactions. 
In the spin sector, the superexchange coupling is $\mib{S}_i \cdot \mib{S}_j$,
isotropic Heisenberg type and independent of the bond direction. 
On the other hand, the superexchange coupling in the orbital sector 
becomes anisotropic because transfer integrals depend on 
orbital indices as well as the bond direction. 
Particularly in the present model (\ref{eq:H_SE^nn}), 
the anisotropy is conspicuous. 
The orbital exchange interaction depends only on the density operator, 
in other words, it is a three--state clock type interaction 
--- there is no quantum fluctuation because the density operator 
is a constant of motion.
Moreover, the orbital interaction depends on the bond direction 
and the orbital states in two sites in a complicated way. 
These features come from the dominant role of 
transfer integrals of $\sigma$ bond, which are orbital diagonal 
and strongly depend on the bond direction and the orbital states.

\subsection{Ground state}
\label{sec:mf-gs}

\begin{figure}[bt]
\centerline{\includegraphics[width=14cm]{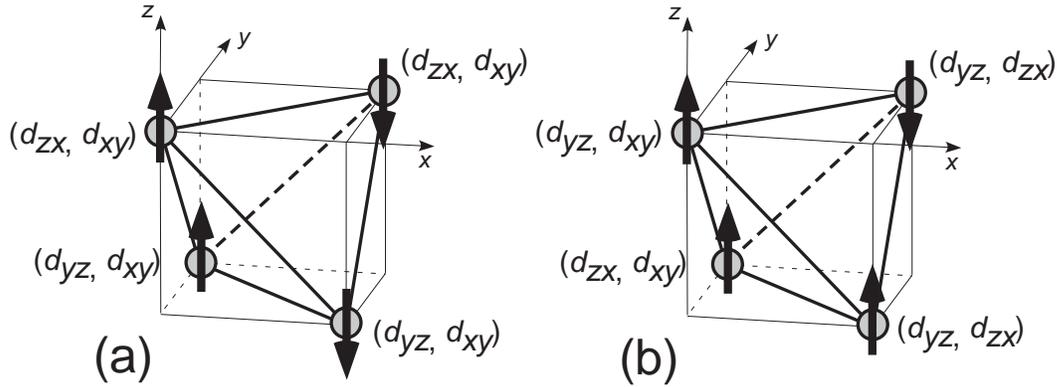}}
\caption{
Candidates for the ground states with four--sublattice ordering 
of the superexchange model (\ref{eq:H_SE^nn}).
Spin and orbital configurations in tetrahedron unit cell are shown.
Arrows denote spins, and labels ($d_{yz}$, $d_{xy}$), etc., 
indicate two occupied orbitals at each site.
}
\label{fig:mf-gs}
\end{figure}

Now we examine the ground state of the superexchange Hamiltonian 
(\ref{eq:H_SE^nn}) defined on the pyrochlore lattice. 
Here, we consider the ground state with 
four--sublattice ordering in V$_4$ tetrahedron unit cell. 
Let us suppose $\eta = 0$ for a while.
In this case, the energy is minimized for 
two different spin--orbital configurations in Fig.~\ref{fig:mf-gs}:
The energy per site is
$E_{\rm SE} = -4J$, 
with counting the interactions with neighboring tetrahedra. 
Per tetrahedron unit cell, 
the configuration (a) contains four spin--antiferro orbital--ferro bonds
(the energy is $-2J$ per bond),
four spin--antiferro orbital--antiferro bonds ($-J$), and
four spin--ferro orbital--antiferro bonds ($-J$). 
The configuration (b) contains four spin--antiferro orbital--ferro bonds,
two spin--antiferro orbital--antiferro bonds, and
six spin--ferro orbital--antiferro bonds.

The degeneracy between these two states is lifted for $\eta > 0$. 
Energies per site for configuration (a) and (b) are given by
\begin{equation}
  E_{\mathrm{SE}}^{\mathrm{(a)}} = -2J(A+C), \quad
  E_{\mathrm{SE}}^{\mathrm{(b)}} = -2J(A+ \frac12 B + C), 
  \label{eq:E_SE}
\end{equation}
respectively. 
Since $B$ is positive, the configuration (b) has a lower energy. 
This appears to be inconsistent with experiments: 
The spin structure in (b) is incompatible with 
the antiferromagnetic state observed in experiments
in Fig.~\ref{fig:neutron} 
because two neighboring spins in the lower $xy$ plane are parallel 
whereas the experimental spin configuration is antiparallel 
in all $xy$ planes.
Moreover, configuration (b) 
noticeably breaks the tetragonal symmetry 
which is experimentally observed in structurally--ordered phase. 
Hence, this mean--field type analysis for the ground state 
tells that the model with superexchange interactions alone 
fails to reproduce the experimental results.
\cite{DiMatteo2005}

There is, however, one caveat ---
the energy difference between the ground state (b) and 
the next lowest state (a) is small compared to the relevant energy scale
in experiments.
The energy difference between (a) and (b) per site is given by
\begin{equation}
\Delta_{\rm SE} = E_{\mathrm{SE}}^{\mathrm{(a)}} - 
E_{\mathrm{SE}}^{\mathrm{(b)}} = JB = \frac{J\eta}{1-3\eta}.
\end{equation}
Let us roughly estimate this energy difference. 
As discussed in Ref.~\citen{Motome2004}, 
estimates for vanadium spinels are 
$J \sim 200$K and $\eta \sim 0.1$.
Hence, the energy difference becomes 
$\Delta_{\rm SE} \sim 30$K, 
which is smaller than the structural transition temperature $\sim 50$K. 
This strongly suggests that spin and orbital configuration 
in the ground state is sensitive to competing interactions; 
in particular, the stable configuration may be changed 
by including the cooperative Jahn--Teller effect, 
which apparently plays an important role in the structural transition. 
Indeed, we will see that the energy for the state (a) can be lower than for (b)
in some parameter regime of the Jahn--Teller coupling 
--- see Sec.~\ref{sec:superexchange+Jahn-Teller}.

\subsection{Instability from high--temperature phase}
\label{sec:mf-highT}

We can gain further insight into the stable spin and orbital configuration 
of the model (\ref{eq:H_SE^nn}) by using another mean--field type argument. 
We here discuss fluctuation effects and an instability 
in spin and orbital sectors 
when we decrease temperature from high--temperature phase. 
\cite{Tsunetsugu2003}

Let us first examine the fluctuations in the spin sector. 
For this purpose, we replace the density operators 
in the orbital part of model (\ref{eq:H_SE^nn})
by their expectation values in the high--temperature regime; 
$n_{i \alpha} \to \langle n_{i \alpha} \rangle = 2/3$. 
Then we end up with the reduced Hamiltonian describing 
the spin sector only, which reads
\begin{equation}
H_{\rm spin}^{\rm nn} = J_{\rm S} \sum_{\langle ij \rangle}
\mib{S}_i \cdot \mib{S}_j,
\label{eq:H_spin^nn}
\end{equation}
up to irrelevant constants, where the effective exchange constant
is given by
\begin{equation}
J_{\rm S} = \frac49 J (C-B).
\end{equation}
Since $J_{\rm S}$ is positive for small $\eta \sim 0.1$, 
the reduced model (\ref{eq:H_spin^nn}) is the antiferromagnetic 
Heisenberg model with $S=1$ spins defined on the pyrochlore lattice.
It is known that the spin correlations hardly 
develop in this model because of the strong geometrical frustration. 
A possibility of a valence--bond--solid type ordering 
with the aid of spin--lattice coupling has been discussed, 
\cite{Yamashita2000}
however it is not easy to explain two successive transitions 
in vanadium spinels by this spin Jahn--Teller scenario. 
The present consideration implies that 
the magnetic instability cannot occur in the high--temperature 
orbital--para phase. 

\begin{figure}[bt]
\centerline{\includegraphics[width=14cm]{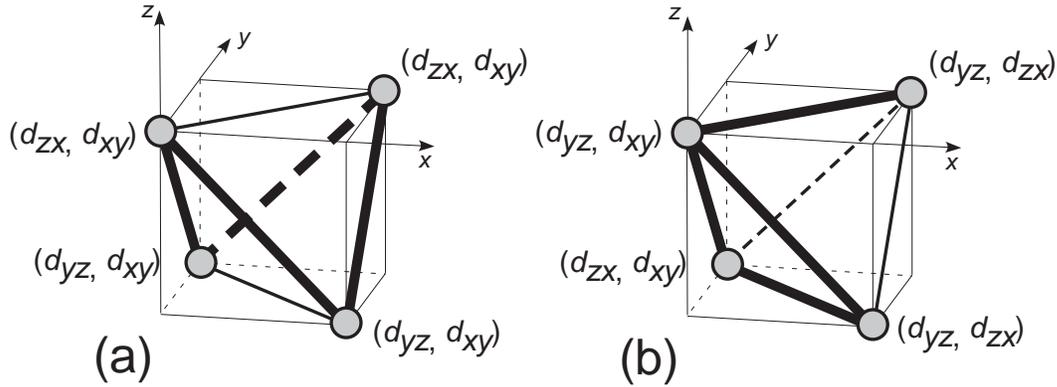}}
\caption{
Stable orbital configurations of the effective orbital model 
(\ref{eq:H_orbital^nn}).
Labels ($d_{yz}$, $d_{xy}$), etc., indicate two occupied orbitals at each site.
Thick lines denote orbital--antiferro bonds.
}
\label{fig:mf-hT}
\end{figure}

Next we consider the instability in the orbital sector. 
In this case, we replace the spin part of model (\ref{eq:H_SE^nn}) 
by their mean value in high--temperature regime, 
that is, $\mib{S}_i \to \langle \mib{S}_i \rangle = 0$. 
The resultant Hamiltonian for the orbital degree of freedom reads
\begin{equation}
H_{\rm orbital}^{\rm nn} = J_{\rm O} \sum_{\langle ij \rangle}
n_{i \alpha(ij)} n_{j \alpha(ij)},
\label{eq:H_orbital^nn}
\end{equation}
up to irrelevant constants, where 
\begin{equation}
J_{\rm O} = J(2A-C).
\end{equation}
This coupling constant $J_{\rm O}$ is also positive so that 
Eq.~(\ref{eq:H_orbital^nn}) is also the antiferro--type model 
on the pyrochlore lattice as the model (\ref{eq:H_spin^nn}). 
A crucial difference between the two models is the form of interactions:
The orbital model (\ref{eq:H_orbital^nn}) has anisotropic 
interaction (three--state clock type) which strongly depends on
the bond direction and the orbital states. 
This peculiar form of the orbital coupling lifts 
the degeneracy in this pyrochlore system, at least partially ---
the three--state clock model still suffers from a frustration
on the pyrochlore lattice. 
The lowest energy is achieved by 
four orbital--antiferro bonds and two orbital--ferro bonds 
within each tetrahedron. 
There are two different types of orbital configurations with these bonds 
as shown in Fig.~\ref{fig:mf-hT}:
In type (a), two orbital--ferro bonds do not touch with each other, 
while they touch at one corner of the tetrahedron in type (b). 
Note that these orbital configurations coincide with 
those in Fig.~\ref{fig:mf-gs}.
The energy of the reduced orbital model (\ref{eq:H_orbital^nn}) 
is the same for these two configurations. 
Hence, there remains degeneracy in the orbital sector 
although the degeneracy is reduced and not severe 
compared to that in the spin sector in the model (\ref{eq:H_spin^nn}).

The degeneracy remaining in the orbital sector 
is easily lifted by including the Jahn--Teller effects. 
This will be discussed in Sec.~\ref{sec:superexchange+Jahn-Teller}.

\section{Model with both superexchange and Jahn--Teller couplings}
\label{sec:superexchange+Jahn-Teller}

In this section, we examine the model with both superexchange interactions 
and Jahn--Teller couplings. 
In Sec.~\ref{sec:mf}, we will show how the mean--field arguments 
in Sec.~\ref{sec:superexchange} are modified by including 
Jahn--Teller distortions in addition to the superexchange model. 
We will point out that once both superexchange and Jahn-Teller effects 
are taken into account, the degeneracy in the orbital sector is lifted 
and an orbital ordering takes place. 
We will also show that spin frustration partially remains 
in the presence of the orbital ordering,
which is finally reduced by farther--neighbor exchange couplings. 
In Sec.~\ref{sec:MC}, the mean--field picture is examined 
and confirmed by Monte Carlo simulation. 
Thermodynamic properties are compared with experimental results. 
In Sec.~\ref{sec:quantum}, effect of quantum fluctuations, 
which is neglected in both mean--field argument and Monte Carlo simulation,
is examined by using linearized spin--wave theory.

\subsection{Mean--field type analysis}
\label{sec:mf}

We have shown in Sec.~\ref{sec:superexchange} that the Kugel--Khomskii type of 
spin--orbital model has two competing stable 
states for V$_4$ tetrahedron unit.  
For realistic values of parameters for the vanadium spinel, 
it was found that the lowest--energy state has a different 
symmetry from the experimental results and 
is not compatible with the antiferromagnetic order in experiments.  
This does not imply the failure of the Kugel--Khomskii 
model but that one needs to take account of couplings 
to lattice distortion.  

Below the structural transition temperature at $\sim 50$K, 
the lattice shows a tetragonal distortion, and 
VO$_6$ octahedra are uniformly compressed along 
the $c$ axis ($c<a$).
Energy level scheme in this case is shown at right most in Fig.~1, and 
three degenerate $t_{2g}$ orbitals in cubic crystal field 
are split to higher--energy doublet and lower--energy 
singlet states in tetragonal field: 
\begin{equation}
  e(d_{yz})=e(d_{zx})= e(t_{2g})+ \frac{\Delta}{3} , \quad
  e(d_{xy})= e(t_{2g})-\frac23 \Delta. 
\end{equation}
As shown in Eq.~(\ref{eq:H_JT^ionic}), the level splitting 
$\Delta = e(d_{yz,zx}) - e(d_{xy})$ is 
proportional to the tetragonal distortion,  
$\Delta = 3 \gamma Q$ with $Q \propto 1- c/a$.  

Let us first reanalyze stable orbital configuration 
in spin paramagnetic phase with taking account 
of the Jahn--Teller effects.  We showed in Sec.~\ref{sec:mf-highT} 
that two types of orbital configurations in Fig,~\ref{fig:mf-hT} 
have the same
energy of the effective orbital interaction (3.12).
Configuration (a) is compatible with the tetragonal symmetry 
as that of the intermediate--temperature phase experimentally observed 
below the structure transition temperature, 
whereas configuration (b) has a lower symmetry.  
Energy gain due to the Jahn--Teller effects is 
largest when the lattice is compressed along the $z$ axis 
for configuration (a) but along the $x$ or $y$ axis 
for configuration (b).  The corresponding energies 
per site are respectively, 
\begin{equation}
  E_{\mathrm{JT}}^{\mathrm{(a)}} = - \frac{\Delta}{3} , \quad 
  E_{\mathrm{JT}}^{\mathrm{(b)}} = - \frac{\Delta}{12} . 
\end{equation}
Therefore, the tetragonal orbital configuration (a) 
has a lower energy and is stabilized by the Jahn--Teller 
effects, as far as a uniform mode of distortion is 
considered.

Next, let us reexamine the stability at $T=0$, 
discussed in Sec.~\ref{sec:mf-gs},
by including the effects of 
tetragonal lattice distortions. 
The energy gain of the configuration is largest 
when the lattice is compressed in the same 
matter as for the previous orbital model.  
The configurations shown in Fig.~2 (a) and (b)
have energies, respectively, 
\begin{eqnarray}
  E_{\mathrm{SE+JT}}^{\mathrm{(a)}} &=& 
  E_{\mathrm{SE}}^{\mathrm{(a)}} + E_{\mathrm{JT}}^{\mathrm{(a)}} =
  -2J(A+C) - \frac{\Delta}{3}, \\ 
  E_{\mathrm{SE+JT}}^{\mathrm{(b)}} &=& 
  E_{\mathrm{SE}}^{\mathrm{(b)}} + E_{\mathrm{JT}}^{\mathrm{(b)}} =
  -2J(A+ \frac12 B + C) - \frac{\Delta}{12}, 
\end{eqnarray} 
where $E_{\mathrm{SE}}^{\mathrm{(a), (b)}}$ are given in Eq.~(\ref{eq:E_SE}). 
Comparing these two, one finds that the configuration (a) 
has a lower energy than (b) when 
\begin{equation}
  \Delta > 4 JB = \frac{4J\eta}{1-3\eta} . 
\end{equation} 
 
As was emphasized in previous sections, 
the vanadium spinels are 
the system in which the Jahn--Teller couplings compete 
with the spin--orbital interactions, 
$\Delta$$\sim$$J\times$(coordination number).  
Recalling that $\eta$ is small, 
we may expect that the above condition is satisfied 
and that the configuration (a) with tetragonal symmetry 
is most stable, if spin--orbital superexchange interactions 
and Jahn--Teller couplings are both considered.
In Sec.~\ref{sec:MC}, we will demonstrate by Monte Carlo calculations 
that the tetragonal orbital order (a) actually appears 
at finite temperatures.   

\begin{figure}[bt]
\centerline{\includegraphics[width=8cm]{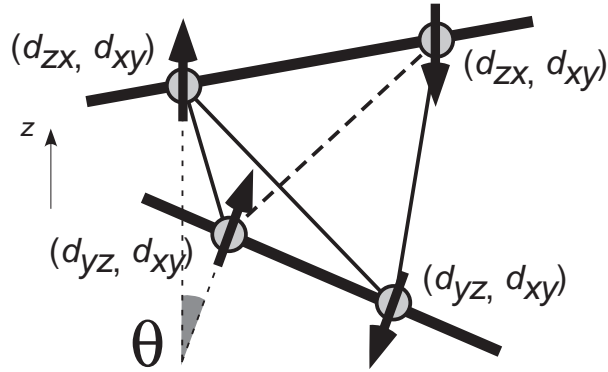}}
\caption{
Frustration between neighboring $xy$ chains 
remaining in the mean--field level.
The mean--field energy does not depend on the relative angle
$\theta$ between the antiferromagnetic chains.
}
\label{fig:nnfrustration}
\end{figure}

Next, let us discuss the spin configuration in the 
magnetic phase.  The configuration in Fig.~\ref{fig:mf-gs}(a)
imposes the spin configuration inside the tetrahedron 
unit cell, and one needs to consider correlations 
between different tetrahedra.  The spin order 
at $\sim 40$K was observed by neutron 
scattering experiment 
\cite{Niziol1973}
and the determined spin pattern is shown in Fig.~\ref{fig:neutron}. 
This pattern corresponds to the $\mib{Q}=(0,0,2\pi /c)$ order
of the configuration in Fig.~\ref{fig:mf-gs}(a) 
($c$ is the length of cubic unit cell along the $z$ axis),
i.e., uniform 
packing in the $xy$ planes and alternate stacking in the $z$ direction 
with spin inversion from layer to layer. 
Antiferromagnetic spin pattern along chains in the $xy$ planes 
are easily explained, 
since any pair of nearest--neighbor spins in $xy$ chains 
should be anti--parallel.  
However, there are no correlations between different $xy$ chains 
within the level of the mean--field approximation.  
Look the spin configuration in Fig.~\ref{fig:mf-gs}(a) 
and rotate two spins in the lower $xy$ plane 
with imposing that they are always anti--parallel 
to each other as in Fig.~\ref{fig:nnfrustration}.  
Such rotations do not change the energy in the mean--field level, 
and this implies no correlation between two chains.  
This is due to the frustrated lattice structure 
and antiferromagnetic nearest--neighbor correlations 
in each chain.

\begin{figure}[bt]
\centerline{\includegraphics[width=14cm]{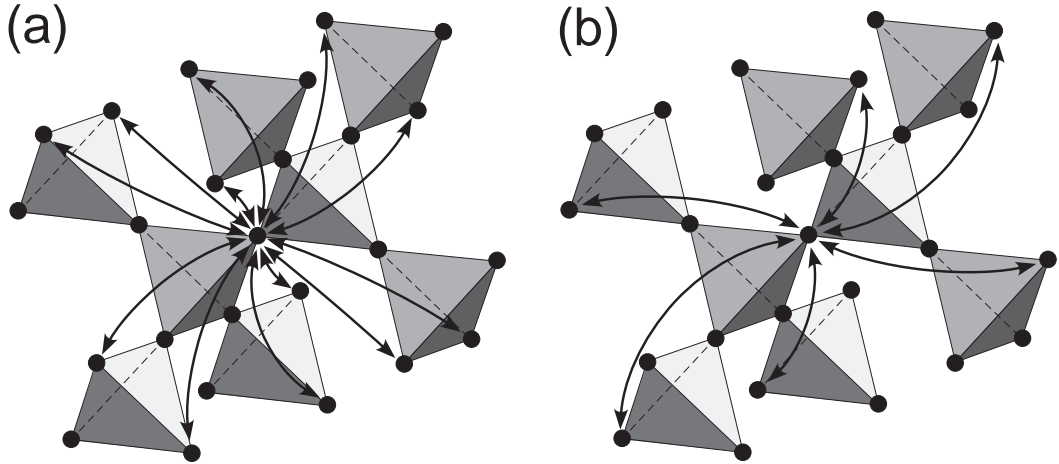}}
\caption{
(a) Second--neighbor exchange couplings, and 
(b) third--neighbor exchange couplings.  
}
\label{fig:J2J3}
\end{figure}

\begin{figure}[bt]
\centerline{\includegraphics[width=14.0cm]{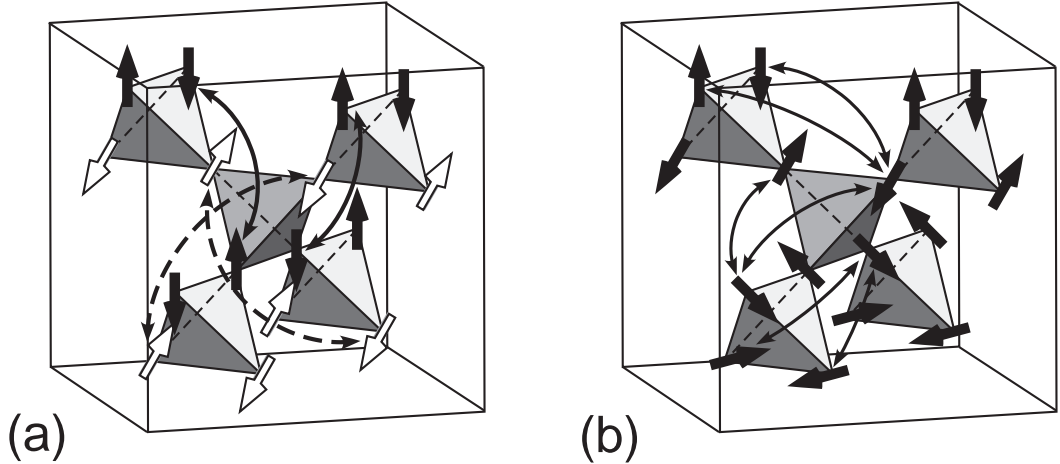}}
\caption{
(a) Two sets of spin chains coupled by third--neighbor 
exchange couplings $J_3$ are depicted in black and white colors.  
(b) Second--neighbor coupling $J_2$ are strongly frustrated 
between neighboring antiferromagnetic chains in the $xy$ planes.  
}
\label{fig:2sublattice}
\end{figure}

Since interchain correlations are not determined by 
nearest--neighbor couplings, we have to consider the 
effects of longer--range couplings, which are much weaker 
in amplitude than the nearest--neighbor ones and are 
usually neglected. These longer--range couplings 
are shown in Fig.~\ref{fig:J2J3}. 
Another possible mechanism is order by disorder.
\cite{Villain1977} 
However, as we will discuss later based on 
numerical (Sec.~\ref{sec:MC}) and 
analytical (Sec.~\ref{sec:quantum}) calculations, 
the mechanism of order by disorder is 
not sufficient to explain a finite static moment in the 
magnetic phase.  
As shown in Fig.~\ref{fig:2sublattice}, 
once antiferromagnetic correlations develop in the $xy$ chains, 
it is important to note that 
third--neighbor couplings lead to more dominant effects 
than second--neighbor ones.  
Third--neighbor couplings connect antiferromagnetic 
chains every two $xy$ planes as shown in Fig.~\ref{fig:2sublattice}(a).  
As a consequence, the 
whole system becomes two interpenetrating sets 
of three--dimensional spin system 
[black and white spins in Fig.~\ref{fig:2sublattice}(a), respectively].  
These two sets 
are connected by nearest--neighbor and second--neighbor couplings.
However, their mean fields are canceled out 
and lead to no effects.  
The frustration of second--neighbor exchanges is depicted 
in Fig.~\ref{fig:2sublattice}(b). 
In this case, it is fluctuations 
which determine the correlation between the two 
sets.  We discussed the effects of quantum fluctuations, 
in particular, in Ref.~\citen{Tsunetsugu2003}, 
and it was found that the collinear 
spin order is stabilized.  Monte Carlo results in the next 
section show that thermal fluctuations also stabilize 
the same collinear order.  

\subsection{Monte Carlo study}
\label{sec:MC}

Now we investigate the thermodynamic properties of 
the model with both superexchange interaction and 
Jahn--Teller coupling by using Monte Carlo method. 
In Sec.~\ref{sec:method}, we briefly describe 
the Monte Carlo technique and the model parameters 
used in the simulation. 
In Sec.~\ref{sec:nn}, the model with nearest--neighbor couplings is studied. 
We will show that the orbital ordering takes place at a finite temperature
but the magnetic frustration remains down to the lowest temperature.
Effects of third--neighbor exchange interactions will be examined 
in Sec.~\ref{sec:3rd}. As predicted in the mean--field argument, 
Monte Carlo results show that the magnetic frustration is reduced
and antiferromagnetic order is stabilized at a finite temperature. 
The phase diagram by controlling the third--neighbor coupling 
will be shown in Sec.~\ref{sec:phasediagram}.

\subsubsection{Method and parameters}
\label{sec:method}

In order to investigate thermodynamic properties of 
the effective spin--orbital--lattice coupled model, 
we use Monte Carlo calculations. 
Quantum Monte Carlo simulation is known to be difficult 
for frustrated models because of the negative sign problem. 
In the present study, we neglect quantum fluctuations 
and approximate the model in the classical level. 
This approximation retains effects of thermal fluctuations 
that may play dominant roles in finite--temperature transitions. 
The quantum nature originates only from spin $S=1$ operators 
because the orbital interaction in Eq.~(\ref{eq:H_SE^nn}) is classical, 
and because Jahn--Teller lattice distortions are treated as 
classical variables in Eq.~(\ref{eq:H_JT}). 
We approximate the spin operators by classical vectors 
with the modulus $|\mib{S}| = 1$ 
to give largest $z$ component $S_z = 1$ in the classical part. 
Therefore, our model for Monte Carlo simulation consists of 
the classical Heisenberg part in the spin sector, 
the three--state clock part in the orbital sector, and 
the classical phonon part. 

We use a standard metropolis algorithm with local updates. 
We typically perform $10^5$ Monte Carlo samplings for measurements 
after $10^5$ steps for equilibration. 
The system sizes in the present work are up to $L=12$, 
where $L$ is the linear dimension of the system measured in the cubic units, 
i.e., the total number of sites $N_{\rm site}$ is up to 
$12^3 \times 16 = 27648$.
We set $J=1$ as an energy unit and the lattice constant of 
cubic unit cell as a length unit. 
We use the convention of the Boltzmann constant $k_{\rm B} = 1$.
The following results are for $\eta = 0.08$, $\gamma^2 = 0.04$ 
and $\lambda = 0.15$. 
Readers are referred to Ref.~\citen{Motome2004} 
for more details of Monte Carlo simulation.

\subsubsection{Model with nearest--neighbor couplings only}
\label{sec:nn}

Let us first examine what happens in the model with 
the nearest--neighbor interactions only, whose Hamiltonian consists of 
the nearest--neighbor superexchange interactions (\ref{eq:H_SE^nn})
and the Jahn--Teller coupling (\ref{eq:H_JT});
\begin{equation}
H = H_{\rm SE}^{\rm nn} + H_{\rm JT}.
\label{eq:H_nn}
\end{equation}
The mean--field type analysis in Sec.~\ref{sec:mf} predicts that 
the orbital--ordered state in Fig.~\ref{fig:mf-hT}(a) is stabilized 
by a cooperative effect of orbital interaction and Jahn--Teller coupling, 
but spin frustration partially remains 
even in the presence of the orbital ordering. 
More precisely, 
the staggered antiferromagnetic correlation develops 
along the chains in the $xy$ planes, 
however the coupling between the neighboring chains remains frustrated: 
Hence, the system cannot develop three--dimensional magnetic 
long--range order. 
We examine this mean--field picture by using Monte Carlo simulation. 

\begin{figure}[bt]
\centerline{\includegraphics[width=9cm]{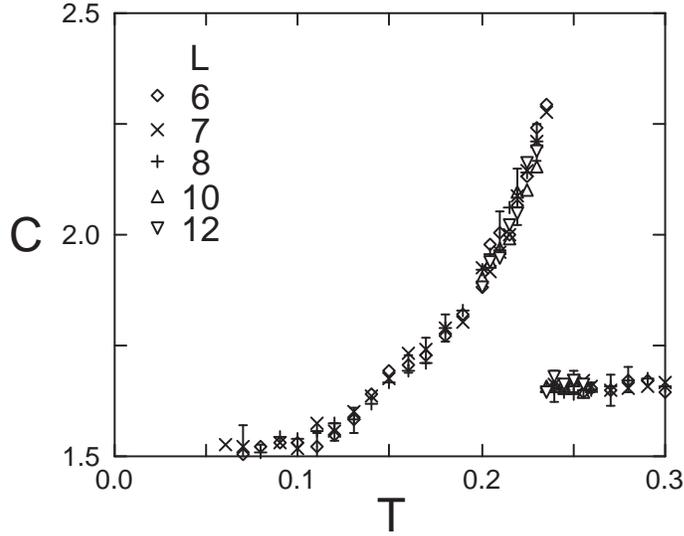}}
\caption{
Temperature dependence of the specific heat per site [Eq.~(\ref{eq:C})]
for the model with nearest--neighbor interactions only. 
}
\label{fig:C_J3=0.0}
\end{figure}

Figure~\ref{fig:C_J3=0.0} shows the temperature dependence of 
the specific heat per site, 
which is calculated by fluctuations of the internal energy as 
\begin{equation}
C = \frac{\langle H^2 \rangle - \langle H \rangle^2}{T^2 N_{\rm site}}.
\label{eq:C}
\end{equation}
The brackets denote thermal averages by Monte Carlo sampling.
The specific heat shows a jump at $T_{\rm O} \simeq 0.23J$, 
indicating a first--order transition. 

\begin{figure}[bt]
\centerline{\includegraphics[width=10cm]{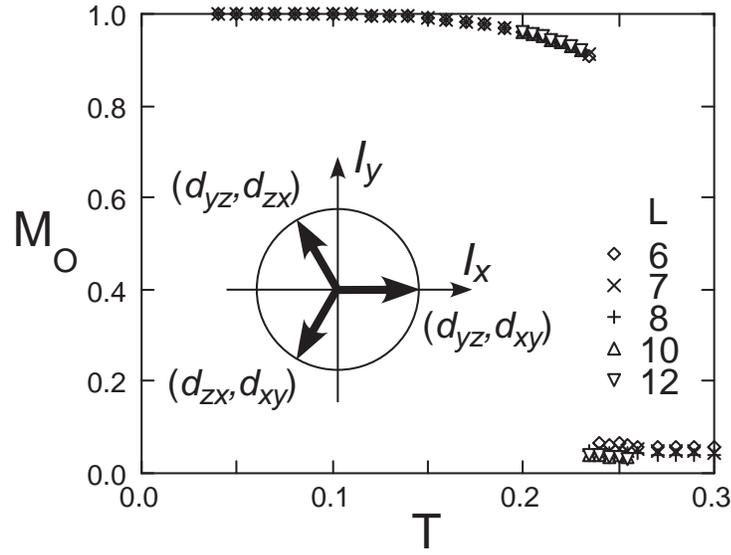}}
\caption{
The sublattice orbital moment in Eq.~(\ref{eq:M_O}). 
The inset shows the three--state clock vector for the orbital state.
}
\label{fig:M_O_J3=0.0}
\end{figure}

At this transition temperature, orbital ordering takes place. 
Figure~\ref{fig:M_O_J3=0.0} shows the sublattice orbital moment, 
which is calculated in the form
\begin{equation}
M_{\rm O} = \frac{4}{N_{\rm site}} 
\Big\langle \Big| \sum_{i \in {\rm sublattice}} \mib{I}_i \Big| \Big\rangle, 
\label{eq:M_O}
\end{equation}
where the summation is taken over the sites within one of 
the four sublattices in tetrahedron unit cell. 
Here, $\mib{I}_i$ is the three--state clock vector at the site $i$ which describes 
three different orbital states as shown in the inset of Fig.~\ref{fig:M_O_J3=0.0}; 
$\mib{I}_i = (1,0)$ for ($d_{yz}, d_{xy}$), 
$\mib{I}_i = (-1/2,\sqrt3/2)$ for ($d_{yz}, d_{zx}$), and 
$\mib{I}_i = (-1/2,-\sqrt3/2)$ for ($d_{zx}, d_{xy}$) 
orbital occupations, respectively. 
It is found that the values of $M_{\rm O}$ for four different sublattices 
have the same value within the error bars 
so that we omit the sublattice index in Eq.~(\ref{eq:M_O}). 
As shown in Fig.~\ref{fig:M_O_J3=0.0}, 
$M_{\rm O}$ shows a clear jump at $T_{\rm O} \simeq 0.23J$ 
where the specific heat jumps. 
This indicates that a four--sublattice orbital ordering occurs at $T_{\rm O}$. 

\begin{figure}[bt]
\centerline{\includegraphics[width=14cm]{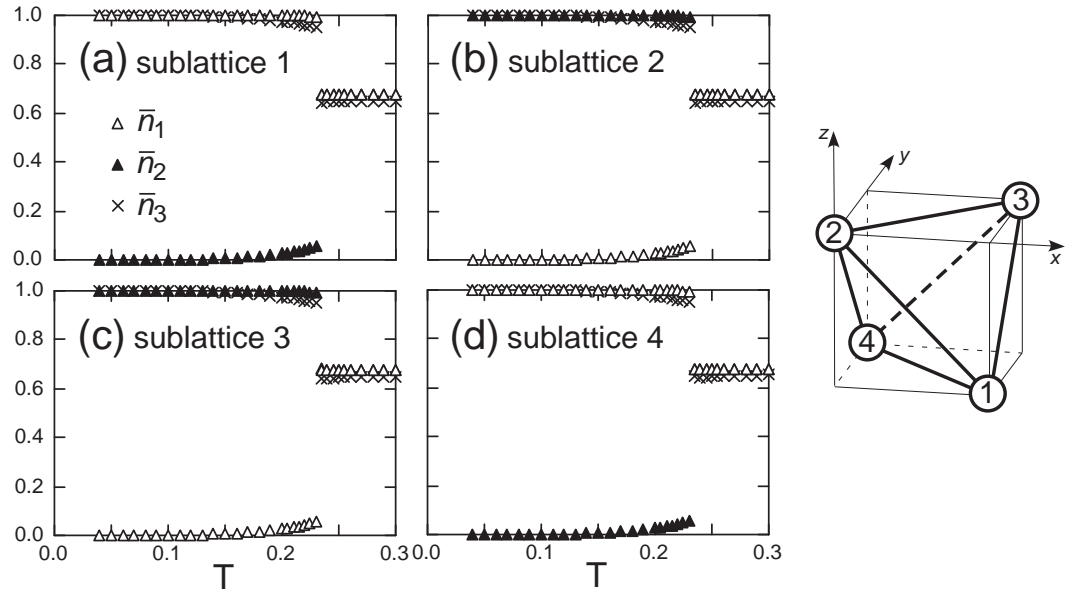}}
\caption{
Electron density in each orbital for four sublattices $1$-$4$ 
defined in Eq.~(\ref{eq:nbar}).
The data are calculated at $L=8$.
}
\label{fig:m_O_distrb_J3=0.0}
\end{figure}

Figure~\ref{fig:m_O_distrb_J3=0.0} plots the orbital distribution 
for four sublattices in tetrahedron unit cell, which is defined as 
\begin{equation}
\bar{n}_\alpha = \frac{4}{N_{\rm site}} \sum_{i \in {\rm sublattice}} 
\langle n_{i \alpha} \rangle, 
\label{eq:nbar}
\end{equation}
where $\alpha = 1$ ($d_{yz}$), $2$ ($d_{zx}$), and $3$ ($d_{xy}$).
The results indicate that 
the orbital distributions suddenly change 
from equally distributed state $\bar{n}_\alpha \sim 2/3$ 
in the high--temperature para phase 
to almost polarized state $\bar{n}_\alpha \sim 0$ or $1$ below $T_{\rm O}$. 
In the orbital ordered phase, 
$d_{yz}$ ($\alpha=1$) and $d_{xy}$ ($\alpha=3$) orbitals are occupied 
in the sublattices 1 and 4, and 
$d_{zx}$ ($\alpha=2$) and $d_{xy}$ ($\alpha=3$) orbitals are occupied 
in the sublattices 2 and 3. 
This orbital ordering structure is consistent with the mean--field prediction 
shown in Fig.~\ref{fig:mf-hT}(a). 

\begin{figure}[bt]
\centerline{\includegraphics[width=9cm]{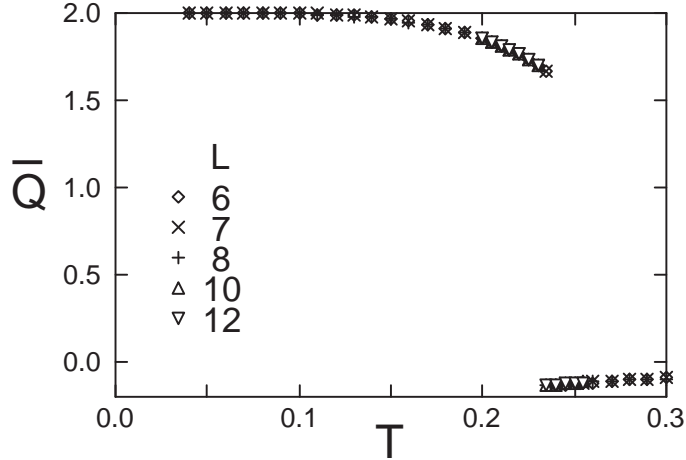}}
\caption{
The average of the Jahn--Teller distortion defined in Eq.~(\ref{eq:Qbar}). 
}
\label{fig:Q_J3=0.0}
\end{figure}

Accompanying the transition at $T_{\rm O}$, 
a tetragonal lattice distortion occurs discontinuously. 
In Fig.~\ref{fig:Q_J3=0.0}, we plot 
the average of the Jahn--Teller distortions which is calculated by 
\begin{equation}
\bar{Q} = \sum_i \frac{\langle Q_i \rangle}{N_{\rm site}}. 
\label{eq:Qbar}
\end{equation}
The positive value of $\bar{Q}$ below $T_{\rm O}$ 
corresponds to a ferro--type tetragonal distortion 
with a compression of VO$_6$ octahedra. 

Therefore, the discontinuous transition at $T_{\rm O} \simeq 0.23J$ is 
ascribed to the orbital ordering with the tetragonal Jahn--Teller distortion. 
The orbital and lattice ordered state in the low--temperature phase 
is consistent with the mean--field prediction.

\begin{figure}[bt]
\centerline{\includegraphics[width=14cm]{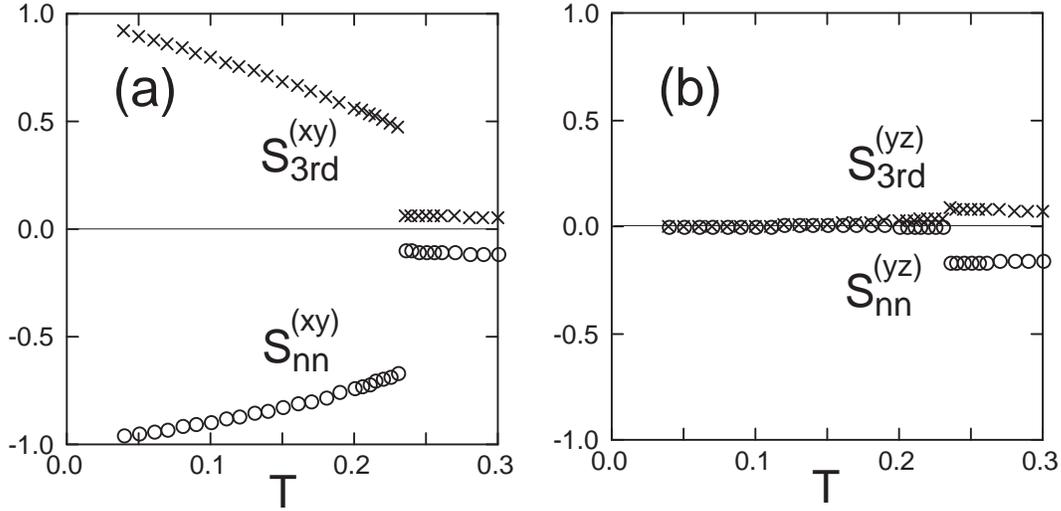}}
\caption{
Spin correlations for the nearest--neighbor sites and 
the third--neighbor sites defined by Eq.~(\ref{eq:sisj})
along the (a) $xy$ chains and (b) $yz$ (or, equivalently $zx$) chains. 
The data are calculated for the model with nearest--neighbor 
interactions only with the system size $L=8$.
}
\label{fig:sisj_J3=0.0}
\end{figure}

So much for the orbital and lattice state --- 
what about the spin sector?
Monte Carlo results clearly indicate that 
spin frustration remains
even below $T_{\rm O}$ as predicted in the mean--field picture. 
Figure~\ref{fig:sisj_J3=0.0} plots temperature dependences of 
spin correlations for nearest--neighbor and 
third--neighbor pairs. 
The spin correlations are measured along the $xy$, $yz$ and $zx$ chains, 
respectively, as
\begin{equation}
S_{\rm nn}^{(\nu)} = \sum_{\langle ij \rangle \in \nu} 
\frac{\langle \mib{S}_i \cdot \mib{S}_j \rangle}{N_{\rm b}}, \quad
S_{\rm 3rd}^{(\nu)} = \sum_{\langle \! \langle ij \rangle \! \rangle \in \nu} 
\frac{\langle \mib{S}_i \cdot \mib{S}_j \rangle}{N_{\rm b}},
\label{eq:sisj}
\end{equation}
where $\nu = xy, yz, zx$, and $N_{\rm b}$ is the number of the bonds 
in the summation. 
The summations with $\langle ij \rangle$ and 
$\langle \! \langle ij \rangle \! \rangle$ are for the nearest--neighbor sites 
and the third--neighbor sites along the $\nu$ chains, respectively.

As shown in Fig.~\ref{fig:sisj_J3=0.0}(a), along the $xy$ chains, 
the antiferromagnetic correlations develop below $T_{\rm O}$. 
On the other hand, the spin correlations are strongly suppressed 
in the $yz$ (and $zx$) directions as shown in Fig.~\ref{fig:sisj_J3=0.0}(b). 
These results support the mean--field picture, 
which predicts the antiferromagnetic correlations along the $xy$ chains 
and strong frustration between the chains. 
Since one--dimensional Heisenberg spin chain cannot order at any temperature 
because of strong fluctuations, 
the model (\ref{eq:H_nn}), which includes only the nearest--neighbor couplings, 
does not show any magnetic ordering at finite temperature. 
This is the reason why we do not see any anomaly 
associated with magnetic ordering in the specific heat 
in Fig.~\ref{fig:C_J3=0.0}.

\subsubsection{Effect of third--neighbor superexchange interaction}
\label{sec:3rd}

The mean--field analysis in Sec.~\ref{sec:mf} predicts that 
the magnetic transition in experiments is reproduced 
by including third--neighbor superexchange interactions. 
Here, we examine this prediction by using Monte Carlo technique. 
Now the total Hamiltonian reads
\begin{equation}
H = H_{\rm SE}^{\rm nn} + H_{\rm SE}^{\rm 3rd} + H_{\rm JT}.
\label{eq:H}
\end{equation}
Here the third--neighbor superexchange term 
$H_{\rm SE}^{\rm 3rd}$ is given by
\begin{equation}
H_{\rm SE}^{\rm 3rd} = -J \sum_{\langle\!\langle ij \rangle\!\rangle} \
[ \ h_{\rm o-AF}^{(ij)} + h_{\rm o-F}^{(ij)} \ ],
\end{equation}
where the summation is taken over the third--neighbor sites 
along the chains, and 
$h_{\rm o-AF(F)}^{(ij)}$ are given 
by Eqs.~(\ref{eq:h_o-AF}) and (\ref{eq:h_o-F}).

\begin{figure}[bt]
\centerline{\includegraphics[width=9cm]{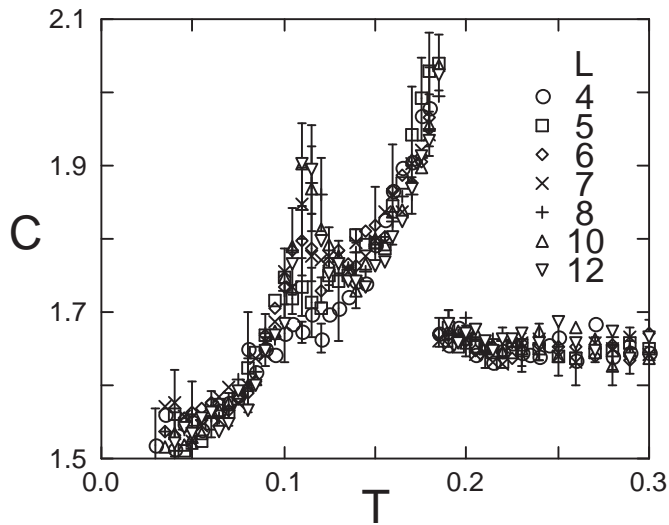}}
\caption{
Temperature dependence of the specific heat per site [Eq.~(\ref{eq:C})]
for the model with the third--neighbor superexchange interaction 
$J_3 = 0.02J$. 
}
\label{fig:C_J3=0.02}
\end{figure}

When we switch on the third--neighbor interaction $J_3$, 
additional anomaly appears in the specific heat 
except for the jump associated with 
the orbital and lattice orderings. 
Figure~\ref{fig:C_J3=0.02} shows the specific heat per site at $J_3 = 0.02J$. 
In addition to the jump at $T_{\rm O} \simeq 0.19J$, 
a peak appears at $T_{\rm N} \simeq 0.115J$. 
This peak, not a jump, increases its height as the system size $L$, 
suggesting a divergence in the thermodynamic limit
--- a characteristic feature of a second--order phase transition. 

The jump at $T_{\rm O} \simeq 0.19J$ is ascribed to 
the discontinuous transition of orbital and lattice orderings. 
The order parameters behave in a similar manner 
as in Figs.~\ref{fig:M_O_J3=0.0} and \ref{fig:Q_J3=0.0} 
although the transition temperature $T_{\rm O}$ slightly decreases. 
On the other hand, the anomaly at $T_{\rm N} \simeq 0.115J$ is ascribed 
to a magnetic ordering as explained in the following. 

\begin{figure}[bt]
\centerline{\includegraphics[width=14cm]{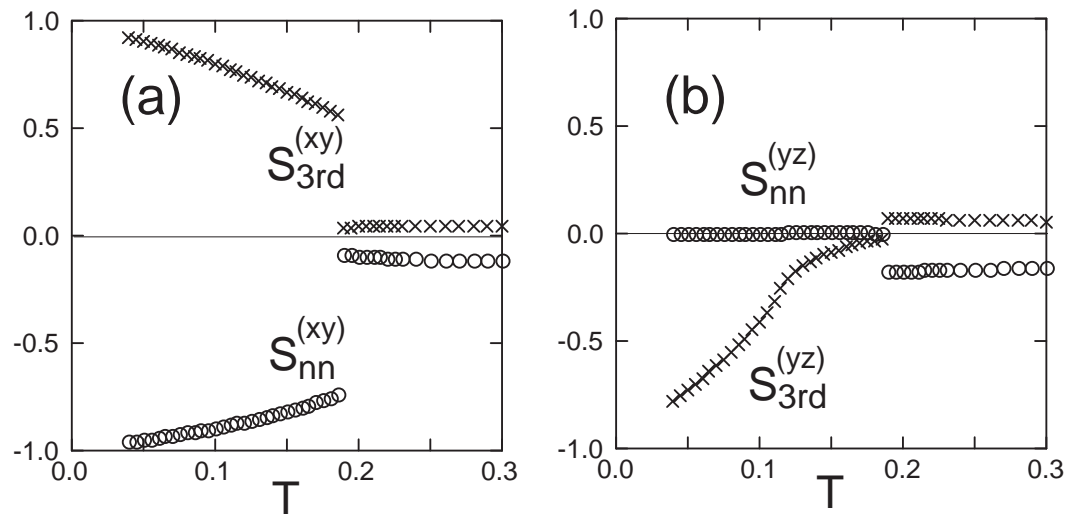}}
\caption{
Spin correlations [Eq.~(\ref{eq:sisj})] for $J_3 = 0.02J$. 
The data are calculated at $L=12$.
}
\label{fig:sisj_J3=0.02}
\end{figure}

Figure~\ref{fig:sisj_J3=0.02} shows the spin correlations 
in Eq.~(\ref{eq:sisj}) for the model (\ref{eq:H}) at $J_3 = 0.02J$. 
Compared to the results at $J_3 = 0$ in Fig.~\ref{fig:sisj_J3=0.0}, 
a big difference is found in the third--neighbor spin correlation 
along the $yz$ (and $zx$) chains; 
the antiferromagnetic correlation rapidly develops 
below $T_{\rm N} \simeq 0.115J$. 
This suggests an emergence of 
three--dimensional magnetic ordering there. 

We calculate the order parameters for the magnetic ordering. 
As described in Sec.~\ref{sec:mf}, in the mean--field level, 
three--dimensional magnetic ordering takes place 
on two independent sublattices which consists of 
($110$) chains and ($1\bar10$) chains, respectively 
[See Fig.~\ref{fig:2sublattice}(a)].
Here, we measure the staggered moment of these sublattice magnetizations, 
which is formally defined by
\begin{equation}
M_{\rm S} = \Big\langle \Big| 
\frac{2}{N_{\rm site}} \sum_i g_i \mib{S}_i \Big|^2 \Big\rangle^\frac12,
\label{eq:M_S}
\end{equation}
where the form factor $g_i$ is given by
\begin{equation}
g_i = \cos[2\pi (x_i + y_i)] + i \cos[2\pi (x_i - y_i)]. 
\end{equation}
Here $x_i$ and $y_i$ are the $x$ and $y$ coordinates of the site $i$ 
measured in the cubic unit, respectively. 
Note that $g_i$ is specified only by the $x$ and $y$ coordinates of the site $i$ 
because the $z$ coordinate is uniquely determined within the cubic unit cell 
due to the special structure of the pyrochlore lattice. 

\begin{figure}[bt]
\centerline{\includegraphics[width=14cm]{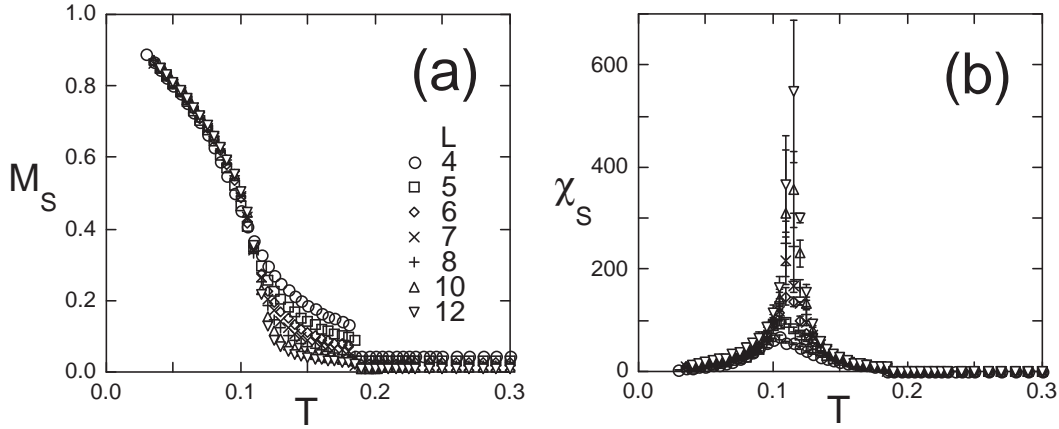}}
\caption{
(a) The staggered moment defined in Eq.~(\ref{eq:M_S}). 
(b) The staggered magnetic susceptibility 
measured by fluctuations of $M_{\rm S}$. 
The results are calculated at $J_3 = 0.02J$.
}
\label{fig:M_S_chi_S}
\end{figure}

Figure~\ref{fig:M_S_chi_S} plots the temperature dependences of 
the sublattice magnetization $M_{\rm S}$ in Eq.~(\ref{eq:M_S}) and 
the corresponding susceptibility $\chi_{\rm S}$. 
As shown in Fig.~\ref{fig:M_S_chi_S}(a), the staggered magnetization 
develops continuously below $T_{\rm N} \simeq 0.115J$ and approaches 
the saturation moment $M_{\rm S} = 1$ as $T \to 0$. 
The susceptibility shows a diverging behavior at $T_{\rm N}$ 
as shown in Fig.~\ref{fig:M_S_chi_S}(b). 
These results indicate that the phase transition at $T_{\rm N}$ is 
a second--order one with the magnetic ordering of staggered sublattice moments. 

In the mean--field level, two sublattice moments are independent. 
It is suggested that quantum fluctuations can align two moments and 
select a collinear magnetic ordering as discussed in Sec.~\ref{sec:mf}. 
While the effect of quantum fluctuations is beyond
our classical approach, 
our Monte Carlo study retains thermal fluctuations 
that are completely neglected in the mean--field argument, and 
it is known that in many frustrated systems 
thermal fluctuations also favor a collinear state. 
\cite{Villain1977}
Here we measure the collinearity 
between two sublattice moments $\mib{M}_1$ and $\mib{M}_2$ by
\begin{equation}
P = \frac32 \Big[ \Big\langle \frac{(\mib{M}_1 \cdot \mib{M}_2)^2}
{(\mib{M}_1)^2 (\mib{M}_2)^2} \Big\rangle -\frac13 \Big]
= \frac32 \Big[ \langle \cos^2 \theta_{12} \rangle -\frac13 \Big], 
\label{eq:P}
\end{equation}
where $\theta_{12}$ is the angle between $\mib{M}_1$ and $\mib{M}_2$.
Note that $P$ becomes $1$ 
when the AF order is collinear, i.e., $\mib{M}_1 \parallel \mib{M}_2$, 
and that $P$ becomes $0$ when $\mib{M}_1$ and $\mib{M}_2$ are independent.

\begin{figure}[bt]
\centerline{\includegraphics[width=9cm]{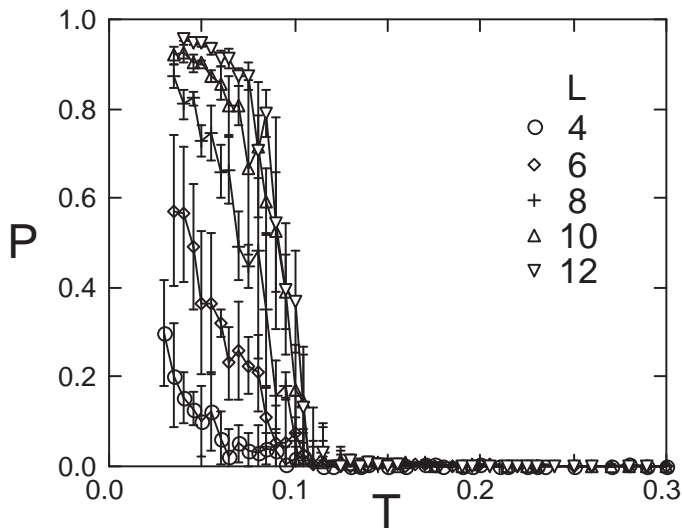}}
\caption{
The measure of collinearity defined in Eq.~(\ref{eq:P}) at $J_3 = 0.02J$. 
The lines are guides for the eyes. 
}
\label{fig:P}
\end{figure}

\begin{figure}[bt]
\centerline{\includegraphics[width=14cm]{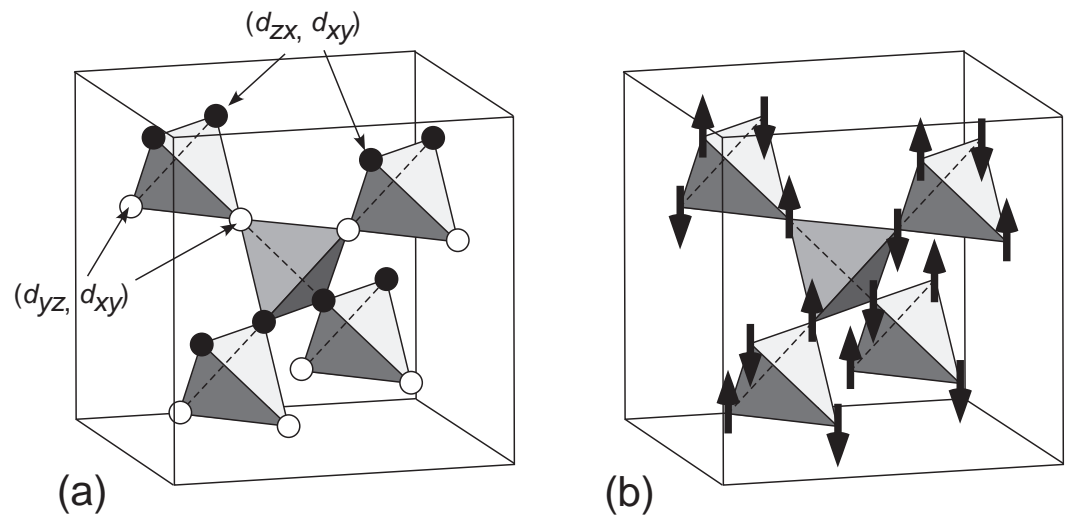}}
\caption{
Monte Carlo results for 
(a) orbital--ordering pattern and (b) spin--ordering pattern 
in the lowest--temperature phase. 
In (a), filled (open) circles denote the sites where 
$d_{zx}$ and $d_{xy}$ ($d_{yz}$ and $d_{xy}$) orbitals are occupied. 
}
\label{fig:so-order}
\end{figure}

Figure~\ref{fig:P} plots the measure of collinearity $P$. 
Below $T_{\rm N} \simeq 0.115J$, $P$ becomes finite and grows rapidly; 
it approaches $1$ as the system size increases.  
This indicates that the magnetic state is collinear below $T_{\rm N}$. 
Consequently, 
the magnetic order is indeed the same one observed in experiments, 
that is, up--down--up--down-- staggered pattern along the $xy$ chains and 
up--up--down--down-- four--times period along the $yz$ and $zx$ chains 
--- see Fig.~\ref{fig:so-order}(b).

Therefore, the present model (\ref{eq:H}), which includes the third--neighbor 
superexchange interactions, successfully explains 
the experimental results for both structural transition and 
antiferromagnetic transition. 
The structural transition at $\sim 50$K is understood as 
the orbital ordering accompanied 
by the tetragonal Jahn--Teller distortion of 
a compression of VO$_6$ octahedra. 
The orbital ordering pattern is a layered type --- the so--called 
A--type order, which is alternative stacking of 
($d_{yz}, d_{xy}$)--occupied layers and 
($d_{zx}, d_{xy}$)--occupied layers in the $z$ direction, 
as shown in Fig.~\ref{fig:so-order}(a). 
The transition at $\sim 40$K is the antiferromagnetic transition. 
The transition temperature $T_{\rm N}$ is lower than $T_{\rm O}$ 
because the spin frustration is reduced 
Below $T_{\rm N}$, the complicated antiferromagnetic ordering 
appears as shown in Fig.~\ref{fig:so-order}(b), 
which is consistent with 
the neutron scattering result in Fig.~\ref{fig:neutron}. 
This magnetic ordering is stabilized by the third--neighbor 
superexchange $J_3$ as well as thermal fluctuations. 
The latter fluctuation effect is important to realize the collinear spin 
configurations by order--by--disorder mechanism.

\subsubsection{phase diagram}
\label{sec:phasediagram}

The antiferromagnetic ordering is stabilized by $J_3$, and hence, 
the transition temperature $T_{\rm N}$ increases as $J_3$. 
On the other hand, the orbital and lattice transition temperature $T_{\rm O}$
slightly decreases as $J_3$ increases 
as found in previous sections. 
What happens for larger $J_3$? 

\begin{figure}[bt]
\centerline{\includegraphics[width=10cm]{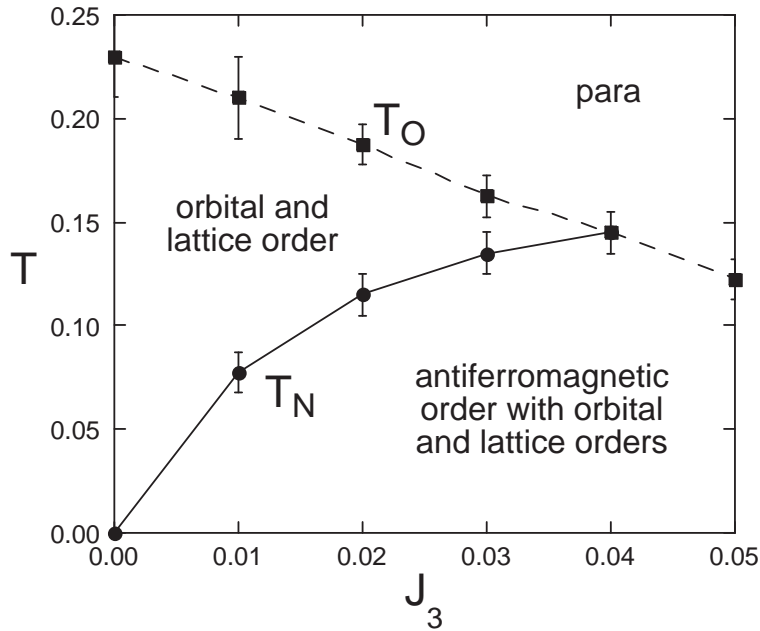}}
\caption{
Phase diagram for the spin--orbital--lattice coupled model 
determined by classical Monte Carlo calculations. 
The orbital and lattice transition at $T_{\rm O}$ (the dashed line) 
is a first--order transition, 
while the antiferromagnetic transition at $T_{\rm N}$ (the solid line) is 
a second--order one. The lines are guides for the eyes. 
}
\label{fig:phasediagram}
\end{figure}

We determined the phase diagram by changing $J_3$. 
The result is shown in Fig.~\ref{fig:phasediagram}. 
The transition temperatures are determined by 
the specific heat, and the orbital, lattice and magnetic order parameters. 
As shown in Fig.~\ref{fig:phasediagram}, 
for large values of $J_3 > 0.03J$, 
$T_{\rm N}$ coincides with $T_{\rm O}$, and there is only 
one discontinuous transition where both orbital and 
antiferromagnetic spin moments becomes finite discontinuously. 

In the region of small $J_3$, our model shows two transitions
at different temperatures as observed in vanadium spinels. 
By using estimates of parameters $J \sim 200$K and 
$J_3 \sim 0.02J \sim 4$K, 
\cite{Motome2004}
the transition temperatures in our results are 
$T_{\rm O} \sim 40$K and $T_{\rm N} \sim 20$K. 
The experimental values are $\sim 50$K and $\sim 40$K, respectively, 
in ZnV$_2$O$_4$. 
The semiquantitative agreement of the transition temperatures is 
satisfactory.

\subsection{Effects of quantum fluctuations}
\label{sec:quantum}

We have determined the ordering pattern of spin configuration 
based on the spin--orbital--lattice model 
and shown that it is 
consistent with the experimental result of elastic 
neutron scattering.  The size of the static magnetic moment, 
however, needs further consideration, since it is subject 
to fluctuations and renormalized to a smaller value. 
To compare with the value at low temperatures in experiment, 
we now show that quantum fluctuations are very large 
in frustrated pyrochlore lattice, 
particularly when the third--neighbor exchange coupling 
$J_3$ is small, and the reduction of the moment is 
prominent. 

\begin{figure}[bt]
\centerline{\includegraphics[height=6cm]{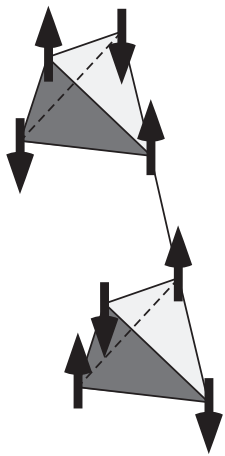}}
\caption{
Magnetic unit cell in the spin ordered phase.  
}
\label{fig:magnetic-uc}
\end{figure}

Let us consider the Heisenberg model,
\begin{equation}
H_{\rm spin} = \sum_{\langle ij \rangle} J_{ij}
\mib{S}_i \cdot \mib{S}_j +
J_3 \sum_{\langle \! \langle ij \rangle \! \rangle}
\mib{S}_i \cdot \mib{S}_j,
\label{eq:H_spin}
\end{equation}
and calculate the moment reduction at zero temperature 
by using the linearized spin--wave theory. 
Here, the first sum is taken over nearest--neighbor pairs, and
the exchange constant is $J_{ij} = J_{xy}$ or $J_{ij} = J_{yz,zx}$ 
depending on the bond direction due to the underlying orbital ordering 
in Fig.~\ref{fig:so-order}(a). 
$J_3$ is the third--neighbor exchange coupling, 
shown in Fig.~\ref{fig:J2J3}(b). 
Note that $J_{xy} \gg J_3$ are both antiferromagnetic,
while $J_{yz,zx}$ is ferromagnetic and $|J_{yz,zx}| \ll J_{xy}$.
Our calculations follow a standard procedure and 
the only special point in the calculations is a frustrated 
geometry of pyrochlore lattice and the anisotropic 
pattern of exchange couplings.  
The starting classical state is the one determined in 
the previous section and shown in Fig.~\ref{fig:so-order}(b).
The magnetic unit cell has eight spins and we choose 
a pair of neighboring two tetrahedra  
(See Fig.~\ref{fig:magnetic-uc}). 
Quantum spin fluctuations around classical 
values are represented by the 
Holstein--Primakoff transformation in terms of boson operators. 
After inserting this for the Heisenberg model (\ref{eq:H_spin}), 
we keep terms up to bilinear in boson operators and 
neglect higher--order terms.  This bilinear boson Hamiltonian 
contains terms of boson--pair creation and annihilation 
and it is diagonalized by a Bogoliubov transformation. 
New boson operators defined in the Bogoliubov transformation 
describe magnetic excitations from the renormalized ground 
state.  The moment reduction is represented by 
energy dispersion of these elementary excitations and 
matrix elements with original bosons introduced 
in the Holstein--Primakoff transformation, 
and is calculated in this way.  

\begin{figure}[bt]
\centerline{\includegraphics[width=9cm]{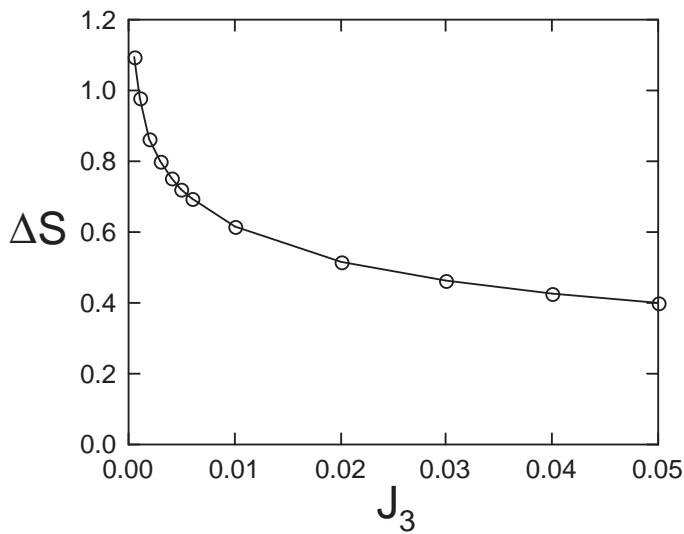}}
\caption{
Moment reduction as a function of $J_3$.
}
\label{fig:deltaS}
\end{figure}

We show the moment reduction $\Delta S$ in Fig.~\ref{fig:deltaS} as
a function of the third--neighbor exchange coupling $J_3$. 
Here, we use 
$J_{xy} = 1$ along the $xy$ chains and
$J_{yz,zx} = -0.1$ along the $yz$ or $zx$ chains.
The renormalized size of static magnetic moment is 
then given as $M_{\rm S} = g \mu_{\rm B} (S-\Delta S)$, where 
$g$ and $\mu_{\rm B}$ are g--factor and Bohr magneton, respectively,
and we set an ordinary value $g=2$, since 
the orbital magnetic moment is now quenched.  
As $J_3$ decreases, $\Delta S$ grows and 
diverges slowly in the $J_3 \rightarrow 0$ limit.  
We show that this divergence is logarithmic 
in the following.  

\begin{figure}[bt]
\centerline{\includegraphics[width=14cm]{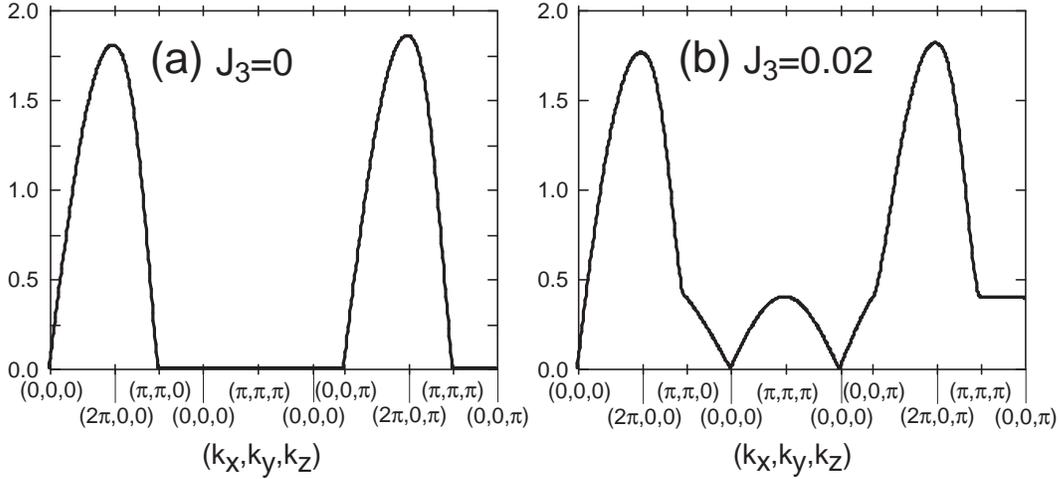}}
\caption{
The lowest magnon mode at (a) $J_3 = 0$ and (b) $J_3 = 0.02$.
}
\label{fig:swdisp}
\end{figure}

An essential point is the presence of zero modes 
at $J_3$=0.  We show the dispersion relation of 
the lowest magnon mode along symmetric axes in the 
Brillouin zone at $J_3 =0$ and $0.02$ in Fig.~\ref{fig:swdisp}. 
The magnon energy vanishes in two planes of 
$k_x = k_y$ and $k_x = - k_y$ as indicated in Fig.~\ref{fig:swdisp}(a). 
Therefore, the magnon 
dispersion is one--dimensional at $J_3 =0$ in this sense. 
As we discussed in a previous section, chains in the $xy$ planes 
have large antiferromagnetic interchain couplings ($J_{xy}$)
and these one--dimensional 
chains are weakly coupled with other chains in adjacent 
$xy$ planes.  Despite the presence of the weak ferromagnetic 
interplane couplings ($J_{yz,zx}$), magnons do not propagate along 
the $z$ direction when $k_x = \pm k_y$.  Each spin in 
$xy$ chain is connected to two spins in a chain in 
an adjacent $xy$ plane.  Quantum interference of the 
two neighbor spins completely suppresses the propagation 
of magnon at these wavevectors, and this is also 
a manifestation of geometrical frustration effects 
of pyrochlore lattice.  In the vicinity of 
$k_x = \pm k_y$ planes, the magnons of the lowest mode 
have a linear dispersion of energy, 
$\omega_{\mib{k}} \propto | \delta \mib{k} |$, 
where $| \delta \mib{k} |$ is the distance 
from the plane of $k_x = \pm k_y$.  
This leads to a logarithmic divergence in the moment 
reduction, $\mathit{\Delta}S \propto \log \mbox{(System Size)}$.  
When the third--neighbor coupling $J_3$ is switched on, 
the zero modes acquire a finite energy that is 
proportional to $J_3$ as shown in Fig.~\ref{fig:swdisp}(b).  
Therefore, the logarithmic 
divergence is cut off at this energy scale, resulting 
in $\mathit{\Delta} S \sim \log (1/J_3)$.
From Fig.~\ref{fig:deltaS}, for realistic value of $J_3 = 0.02J$, 
the moment reduction becomes $\mathit{\Delta}S \sim 0.5$. 
This leads to a largely reduced staggered moment
$M_{\rm S} \sim 1$[$\mu_{\rm B}$], 
which is consistent with experimental results.  
\cite{Lee2004}

\section{Discussion}
\label{sec:disc}

\subsection{Effect of relativistic spin--orbit coupling}
\label{sec:LS}

Thus far, we neglected relativistic spin--orbit coupling,
and employed the model that describes an interplay  
between the spin--orbital superexchange interactions 
and 
the cooperative Jahn--Teller coupling of tetragonal symmetry. 
We showed that the model successfully explains many experimental results 
of thermodynamic properties in vanadium spinel oxides. 
After our works, there have been complemental studies 
which theoretically examine effects of the relativistic spin--orbit coupling.
Here, we briefly discuss about them 
in comparison with our work. 

Tchernyshyov proposed a model assuming the dominant role of 
the relativistic spin--orbit coupling. 
\cite{Tchernyshyov2004}
In his model, an effective angular momentum $\mib{L}'$ 
with the magnitude $L'=1$ 
is introduced to describe ($t_{2g}$)$^2$ configurations. 
The orbital angular momentum is given by $\mib{L} = - \mib{L}'$, and 
couples to the $S=1$ spin $\mib{S}$ at each site antiferromagnetically, 
and consequently, 
the total momentum $\mib{J}' = \mib{L}' + \mib{S}$ 
with the magnitude $J'=2$ is a conserved, 
good quantum number in the system. 
The lowest energy state is the $J' = 2$ quintuplet. 
When the Jahn--Teller effect is included as a perturbation,
a degeneracy remains partially in this manifold: 
The ground state energy is the same between 
$J'_z = 0$ state with an elongation of tetrahedron and
$J'_z = \pm 2$ states with a compression of tetrahedron. 
Moreover, this degeneracy cannot be lifted even when 
the superexchange interactions are included. 
Consequently, the model fails to reproduce by itself 
the spin and orbital ordered state observed in experiments. 
By putting as an input the experimental fact that 
the magnetic ordering takes place, 
the model predicts the ground state with $J'_z = \pm 2$ and 
a compression of tetrahedron. 

The orbital ordering predicted in this way is different from our result. 
In Tchernyshyov's model, 
the orbital occupation becomes uniform in space, that is, at all sites
\begin{equation}
\langle n_{i1} \rangle = \langle n_{i2} \rangle = \frac12, \quad
\langle n_{i3} \rangle = 1.
\label{eq:uniform-oo}
\end{equation}
The orbital state is a complex one, 
$| d_{yz} \pm i d_{zx} \rangle$ at all sites. 
On the other hand, our model predicts the A--type orbital ordering 
with a staggered occupation of $d_{yz}$ and $d_{zx}$ 
orbitals in the $z$ direction as shown in Fig.~\ref{fig:so-order}(a). 
As pointed out in Ref.~\citen{Tchernyshyov2004}, 
the crystal symmetry is different between these two orbital states. 
The A--type ordering in Fig.~\ref{fig:so-order}(a) breaks 
mirror reflections in ($110$) and ($1\bar10$) planes as well as 
diamond glides in ($100$) and ($010$) planes, 
on the other hand, the uniform ordering in Eq.~(\ref{eq:uniform-oo})
preserves these symmetries. 

Experimentally, the lattice symmetry below the structural transition 
temperature has not been completely determined yet. 
X--ray scattering results for polycrystal samples 
suggest the persistence of the mirror and glide symmetry 
at the low temperature phase. 
\cite{Nishiguchi2002,Reehuis2003} 
On the contrary, recent synchrotron X--ray data 
for a single crystal sample suggest a breaking of the glide symmetry 
although they cannot decide whether the mirror symmetry is broken or not. 
\cite{Lee2004} 
This controversy probably comes from a small electron--phonon coupling 
in this $t_{2g}$ system. 
In order to settle the theoretical controversy,  
sophisticated experiments with large single crystals are highly desired. 

Another model has been proposed by Di Matteo {\it et al.}
\cite{DiMatteo2005}
The model takes into account 
the spin--orbital superexchange interaction and 
the relativistic spin--orbit coupling on an equal footing. 
Mean--field analysis was applied to obtain the ground state. 
In the absence of the relativistic spin--orbit coupling, 
the ground state is the one shown in Fig.~\ref{fig:mf-gs}(b) 
for finite Hund's--rule coupling $\eta > 0$ 
as discussed in Sec.~\ref{sec:mf-gs}. 
When the relativistic spin--orbit coupling is turned on, 
the orbital state changes into a highly antisymmetric one 
(four sites in tetrahedron unit cell take different 
orbital states including both real and complex ones) 
with keeping the 3--up and 1--down spin configuration. 
Finally, for moderate values of the relativistic spin--orbit coupling, 
the ground state shows a transition into 
the spin and orbital ordered state 
which is compatible with Tchernyshyov's prediction 
by assuming the dominant relativistic spin--orbit coupling. 
In this sense, this study may interpolate our model and 
the Tchernyshyov's model. 
In their treatment, however, the Jahn--Teller coupling is 
completely neglected. 
As we have seen in Sec.~\ref{sec:mf}, 
the mean--field ground states are drastically changed
by the effects of the Jahn--Teller lattice distortion. 
It is interesting to investigate with including 
the cooperative aspect of the lattice distortions.

\subsection{Direction of magnetic moment}
\label{sec:dir}

Experimentally, the spontaneous magnetic moment 
appears along the $z$ direction. 
This is naturally explained by the scenario based on 
the relativistic spin--orbit coupling in the previous section as follows. 
The complex orbital state $| d_{yz} \pm i d_{zx} \rangle$ 
has a finite matrix element of the $z$ component of 
orbital angular momentum $\mib{L}$. 
This aligns the magnetic moment along the $z$ direction 
through the spin--orbit coupling. 
On the other hand, the spin part of our model is completely isotropic, 
and no direction is preferred for spin ordering.  
Therefore, one needs to introduce a process 
with spin anisotropy to explain the observed spin direction.  
Here, we discuss that the direction of moment is reproduced 
by our model when we include quantum fluctuations in the orbital sector 
as well as the relativistic spin--orbit coupling 
in a perturbative way. 

As described in Sec.~\ref{sec:SEmodel}, 
the orbital part of our model is completely classical, 
three--state clock type interaction, 
because we take into account the dominant $\sigma$-bond transfers only. 
If we include small contributions from 
$\pi$ and $\delta$-type transfer integrals, 
the orbital interaction is no longer classical, 
that is, contains small off--diagonal exchange interactions. 
This should lead to quantum fluctuations in the orbital sector. 
Our results in Sec.~\ref{sec:superexchange+Jahn-Teller} showed that 
$d_{xy}$ orbital is lowered and occupied at every site, and that
$d_{yz}$ and $d_{zx}$ orbitals are alternatively occupied 
as long as the $\sigma$ contribution is dominant. 
Under this situation, we expect that the orbital quantum nature, 
coming from small $\pi$ and $\delta$ contributions, 
leads to fluctuations between $d_{yz}$ and $d_{zx}$ orbitals dominantly. 
This $d_{yz}$--$d_{zx}$ fluctuation gives rise to 
a finite matrix element of $L_z$. 
As mentioned above, 
$\mib{L}$ couples to the spin $\mib{S}$ antiferromagnetically 
once we consider a finite relativistic spin--orbit coupling, 
and therefore, the induced $L_z$ component aligns 
the magnetic moment in the $z$ direction.

\subsection{Weakly-coupled $S=1$ chains --- Haldane gap?}
\label{sec:Haldane}

As discussed in Sec.~\ref{sec:superexchange+Jahn-Teller}, 
the orbital ordering reduces the magnetic frustration in our model. 
Below the orbital ordering temperature, 
antiferromagnetic correlations develop along the $xy$ chains, and 
the coupling is frustrated between neighboring chains. 
Hence, the system can be regarded as
weakly--coupled $S=1$ spin chains under orbital ordering. 
It is well known that the ground state of antiferromagnetic $S=1$ chain 
is spin--singlet, and a gap opens between the ground state 
and the lowest triplet excitation --- the so--called Haldane gap. 
\cite{Haldane1983}
This is purely quantum mechanical effect, and hence 
it is beyond our present theory by mean--field argument 
and Monte Carlo simulation
to examine the possibility of Haldane gap. 
Experimentally, vanadium oxides exhibit 
magnetically-ordered ground state, not the spin-singlet state, and
we here address this issue in the light of our results 
and make some remarks. 

It is known that the Haldane gap is fragile against the interchain couplings. 
For example, when antiferromagnetic $S=1$ chains are connected 
to form of a simple cubic lattice structure, 
the critical value of the interchain exchange coupling 
is estimated at $2 - 3$\% of the intrachain exchange coupling.
\cite{Koga2000} 
In our pyrochlore system, the relevant interchain coupling is 
the third--neighbor exchange $J_3$ in Fig.~\ref{fig:J2J3}(a).
The coupling $J_3$ constitutes a complicated three-dimensional network 
without any frustration among the chains. 
In addition, the neighboring $xy$ chains are coupled 
by weak ferromagnetic coupling as well as 
second--neighbor coupling 
although they are frustrated due to the pyrochlore geometry. 
We expect for realistic parameter regime
that the Haldane gap state is unstable 
against these interchain couplings. 

On the other hand, as shown in the phase diagram in Fig.~\ref{fig:phasediagram}, 
the transition temperature for the magnetic ordering 
decreases rapidly as $J_3$, and becomes zero in the limit of $J_3$
within our classical model. 
This implies that once all farther--neighbor couplings are irrelevant,
a competition occurs between the long-range magnetic order and 
the quantum Haldane gap formation, 
leading to a quantum critical point.  
This interesting problem is left for further study.

\section{Summary}
\label{sec:summary}

In this paper we have reviewed our recent studies on the
geometrically
frustrated systems, vanadium spinel oxides $A$V$_2$O$_4$ 
with divalent $A$--site cations.  
These are spin--1 antiferromagnets on the
frustrated pyrochlore network of vanadium sites.
We have employed an exchange model of electron spin and
orbital degrees of freedom with including coupling
to Jahn--Teller lattice distortion.
By using mean--field approach and numerical calculations,
we have shown that this model can describe two
phase transitions in this system.  The structural transition
at a higher temperature is ascribed to orbital ordering
accompanied by lattice distortion.  The orbital ordering
leads to an spatially anisotropic pattern of spin exchange
couplings. Geometrical frustration of spins is thus
relaxed, and this induces a magnetic transition.
Spin configuration in the magnetic phase was determined
and we found that it agrees with the one observed
by neutron scattering experiments.  It was also found
that the third--neighbor spin exchange couplings are
important to determine the spin configuration and
the magnetic transition temperature.  A realistic
value of the third--neighbor coupling gives a size
of the magnetic moment and a transition temperature,
both of which are quite well in agreement with the
experimental values.  These semiquantitatively good
agreements with experimental results support our choice
of model Hamiltonian and also our picture of the two
transitions.  A few points, however, remain to be
investigated in more details.  One is more
systematic analysis of the Jahn--Teller effects.
So far, we have considered only uniform modes of lattice
distortion, experimentally observed ones, but
it is important to check that the uniform mode
is the most unstable one.  Another point is more
quantitative analysis of the effects of transverse
orbital fluctuations.  These are introduced by electron
hoppings in $\pi$-- and $\delta$--bonds.
The orbital part of the effective exchange model
remains strongly anisotropic of three--state Potts
type even with including transverse orbital
couplings, but it is interesting to investigate
the effects of these couplings.  The transverse
couplings induce orbital angular momentum,
and therefore the relativistic spin--orbit coupling
may show interesting effects.

\section*{Acknowledgements}
We would like to thank M. Onoda and S.-H. Lee for fruitful discussions. 
Y. M. would like to thank D.I. Khomskii, G. Jackeli and C. Yasuda 
for helpful comments. 
This work was supported under a Grant--in--Aid for Scientific Research
(No. 16GS50219) and NAREGI 
from the Ministry of Education, Science, Sports, and Culture of Japan

%

\end{document}